\documentclass[aps,preprint,amsmath,amssymb,t1,11pt]{revtex4}
\usepackage{graphicx}
\usepackage{epstopdf}
\usepackage{xcolor}
\usepackage{slashed}
\usepackage{multirow}
\usepackage{booktabs} 
\usepackage{adjustbox}
\usepackage{rotating}
\usepackage[colorlinks=true,
            linkcolor=black,
            citecolor=blue,
            urlcolor=black]{hyperref}

\newcommand{\proaa}{\mu^{+}\mu^{-}\to\mu^+\gamma\gamma\mu^-}
\newcommand{\proza}{\mu^{+}\mu^{-}\to\mu^+Z\gamma\mu^-}
\newcommand{\com}{\sqrt{s}}
\newcommand{\couplings}{$f_{T,i}/\Lambda^{4}$ (i=0,..,9) }

\newcommand{\madgraph}{\texttt{MadGraph5\_aMC@NLO} }
\newcommand{\feynrules}{\texttt{FEYNRULES} }
\newcommand{\ufo}{\texttt{Universal FeynRules Output (UFO)} }
\newcommand{\delphes}{\texttt{Delphes} }
\newcommand{\pythia}{\texttt{Pythia8} }

\begin{document}
\draft

\title{Constraints on Anomalous Quartic Gauge Couplings via $\gamma\gamma$ and $Z\gamma$ Vector Boson Scattering at Muon Colliders}
\author{M. Tekin}
\email[]{wustafaatekin@gmail.com}
\affiliation{Department of Physics, Bolu Abant Izzet Baysal University, 14280, Bolu, T\"{u}rkiye.}

\author{A. Senol}
\email[]{senol\_a@ibu.edu.tr}
\affiliation{Department of Physics, Bolu Abant Izzet Baysal University, 14280, Bolu, T\"{u}rkiye.}

\author{H. Denizli}
\email[]{denizli\_h@ibu.edu.tr}
\affiliation{Department of Physics, Bolu Abant Izzet Baysal University, 14280, Bolu, T\"{u}rkiye.}

\date{\today}

\begin{abstract}

In the Standard Model, the couplings between gauge bosons are tightly constrained by the principles of gauge symmetry and renormalizability. However, the presence of anomalous couplings suggests the possibility of new physics beyond the Standard Model (BSM).
In this study, we focus on the sensitivities of anomalous quartic gauge couplings (aQGCs), specially the dimension–8 operators associated with field–strength tensor structures within the effective field theory (EFT) framework, at future Muon Colliders. Our analysis targets the neutral aQGC-sensitive processes $\proaa$ and $\proza$, simulated at center-of-mass energies of 3~TeV and 10~TeV. Signal and background events are generated using $\madgraph$, interfaced with $\pythia$ for parton showering and hadronization, and $\delphes$ for fast detector simulation. A multivariate analysis based on Boosted Decision Trees (BDTs) is employed to enhance signal-to-background discrimination, utilizing a comprehensive set of kinematic and reconstructed observables from the final-state particles. Unitarity is preserved through the application of an energy–dependent clipping procedure within the EFT validity regime.

Our findings indicate that future muon colliders offer significant sensitivity improvements over current experimental constraints on aQGCs. Furthermore, a comparison with other future collider scenarios shows that the 10 TeV Muon Collider, even with a 10\% systematic uncertainty, provides substantially stronger projected limits at 95\% confidence level than those currently reported by the ATLAS collaboration at the LHC as well as projected limits by future hadron colliders. These results underscore the enhanced potential of high-energy muon collider to probe new physics in the electroweak sector through precision measurements of aQGCs.
\end{abstract}


\maketitle

\section{Introduction}

The non-Abelian gauge structure of the electroweak (EW) sector within the Standard Model (SM) provides a unique laboratory for probing the self-interactions of gauge bosons. In particular, quartic gauge couplings (QGCs) are directly linked to the mechanism of electroweak symmetry breaking (EWSB) and are therefore highly sensitive to possible effects of physics beyond the Standard Model (BSM). The SM predicts quartic vertices involving charged gauge bosons at the tree level, such as $WWVV$ where $V=W^{\pm},\gamma,Z$. In contrast, purely neutral quartic gauge interactions of the form $V'V'V'V'$ ($V'=\gamma,Z$) are absent at tree level and arise only through higher-order loop corrections. Consequently, any measurable deviation in these neutral interactions would constitute a clear signature of new physics.

Such loop-suppressed neutral interactions provide an excellent window for probing deviations induced by heavy new physics. A model-independent description of such potential deviations can be formulated within the Effective Field Theory (EFT) framework, where the SM Lagrangian is extended by higher-dimensional operators constructed from SM fields~\cite{Buchmuller:1985jz, Eboli:2006wa}. In this approach, the lowest-order operators capable of modifying QGCs without simultaneously inducing anomalous triple gauge couplings arise at dimension-8 and can be expressed as
\begin{equation}
\mathcal{L}_{\text{EFT}} = \mathcal{L}_{\text{SM}} + \sum_{i}\frac{f_i}{\Lambda^4}\mathcal{O}_i^{(8)},
\end{equation}
where $f_i$ denote the corresponding Wilson coefficients and $\Lambda$ represents the energy scale of new physics.

Experimentally, vector boson scattering (VBS) processes provide a powerful probe of these operators. Although an extensive VBS program has been carried out at the Large Hadron Collider (LHC) and its High-Luminosity upgrade, the sensitivity to neutral quartic interactions remains experimentally challenging due to overwhelming QCD-induced backgrounds and associated systematic uncertainties. Lepton colliders offer a substantially cleaner experimental environment; however, proposed electron--positron facilities typically face a trade-off between achievable center-of-mass energy and integrated luminosity.

In this context, a high-energy muon collider provides a qualitatively new opportunity to explore anomalous neutral quartic gauge couplings at unprecedented energy scales. Owing to the fundamental nature of muons, the full center-of-mass energy is available for the hard interaction, in contrast to hadron colliders where the effective collision energy is reduced by parton distribution functions. Moreover, at multi-TeV energies, muons effectively act as sources of collinear electroweak gauge bosons, enabling the $\mu^+\mu^-$ collider to operate as a vector boson collider~\cite{Accettura:2023ked,Aime:2022flm}. Since the EFT-induced contributions to aQGC processes exhibit a strong energy dependence, typically scaling as $\sigma_{\mathrm{aQGC}}\sim s^{n}/\Lambda^{4}$, the high-energy reach of a muon collider is expected to significantly extend the discovery potential beyond current LHC limits~\cite{MuonCollider:2022xlm, MuonCollider:2022nsa}.

Future muon colliders provide a unique environment for probing physics beyond the Standard Model, particularly through precision tests of anomalous quartic gauge couplings (aQGCs). Recent studies have demonstrated the high sensitivity of muon colliders to dimension-8 operators via processes such as $\mu^+\mu^-\to \bar{\nu}\nu\gamma\gamma$~\cite{Chen:2025mxf}, triboson production~\cite{Yang:2020rjt}, and vector boson scattering processes including $W^+W^- \to W^+W^-$~\cite{Yang:2022fhw}. In particular, the investigation of neutral quartic gauge couplings such as $Z\gamma\gamma\gamma$, $ZZ\gamma\gamma$, and $ZZZ\gamma$ plays a crucial role in understanding possible deviations within the electroweak sector~\cite{Gutierrez-Rodriguez:2025wcy}. Moreover, the phenomenology of gluonic quartic gauge couplings at muon colliders has also received considerable attention in the literature~\cite{Guo:2025pht, Yang:2023gos}. From the data analysis perspective, recent efforts have gone beyond traditional cut-based approaches by incorporating machine learning techniques such as Artificial Neural Networks (ANN)~\cite{Yang:2022fhw}, Principal Component Analysis (PCA)~\cite{Dong:2023nir}, K-means clustering algorithms~\cite{Zhang:2023yfg, Zhang:2024ebl}, and the nested local outlier factor method~\cite{Chen:2025mxf}, which significantly improve signal sensitivity and background rejection efficiency~\cite{Zhang:2023yfg, Dong:2023nir}. It is worth noting how our study complements and extends existing investigations of anomalous quartic gauge couplings (aQGCs) at future muon colliders. In Ref.~\cite{Amarkhail:2023xsc}, the $\gamma\gamma\gamma \gamma$ vertex is parameterized in terms of two effective couplings $\zeta_1$ and $\zeta_2$, and the sensitivity of the $\mu^+\mu^- \to \mu^+\gamma\gamma\mu^-$ process is evaluated for several center-of-mass energies. A subsequent work, Ref.~\cite{Amarkhail:2024kfq}, explores the anomalous $\gamma\gamma\gamma Z$ vertex at $\mu^+\mu^-$ colliders; again, detector simulation and full event generation are not included while these works provide valuable insight into a specific neutral quartic interaction, it adopts a vertex-based effective description distinct from the dimension‐8 EFT framework used here, and does not include realistic detector effects or parton showering in the signal and background modeling. Similarly, Ref.~\cite{Gutierrez-Rodriguez:2025wcy} employs the same dimension‐8 $f_{T,j}/\Lambda^4$ notation as our analysis to study the $\mu^+\mu^- \to \mu^+\mu^- Z\gamma$ process and projects sensitivities at $\sqrt{s}=10\,$TeV. However, that work implements only parton‐level computations without showering or detector simulation, leaving out important experimental effects that can impact event selection and signal discrimination. In contrast, 
we investigate the sensitivity of a future muon collider to dimension-8 anomalous quartic gauge couplings through the neutral VBS processes
\begin{equation}
\proaa \quad \text{and} \quad 
\proza
\end{equation}
Two benchmark energy stages are considered in accordance with the International Muon Collider Collaboration design study~\cite{MuonCollider:2022xlm,InternationalMuonCollider:2024jyv}: a $\sqrt{s}=3$ TeV configuration with an integrated luminosity of $1~\mathrm{ab}^{-1}$, and a $\sqrt{s}=10$ TeV stage targeting $10~\mathrm{ab}^{-1}$. To enhance the separation between the rare EFT-induced signal and the dominant Standard Model background contributions, a multivariate analysis based on the Boosted Decision Tree (BDT) algorithm within the TMVA framework is employed.

The remainder of this paper is organized as follows. Section~\ref{secII} introduces the EFT formalism and the relevant dimension-8 operators. Section~\ref{secIII} describes the event generation, detector simulation, and multivariate analysis strategy. Section~\ref{secIV} presents the projected sensitivity limits for the considered collider stages and compares our results with current experimental bounds. Section~\ref{conc} summarizes the work and discusses the implications of our findings.

\section{Dimension-8 anomalous Quartic Gauge Couplings in EFT Framework}\label{secII}
\begin{figure}[h!]
\includegraphics[scale=0.90]{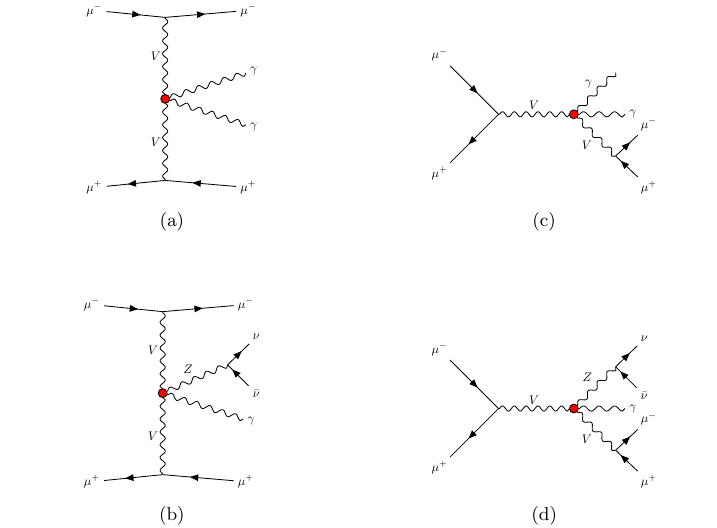}
\caption{Feynman diagrams illustrating the contribution of aQGC to (a) and (b) the VBS processes , (c) and (d) the non-VBS processes in $\mu^+\gamma\gamma\mu^-$ and $\mu^+Z\gamma\mu^-$ productions at a Muon Collider.\label{fd}}
\end{figure}
In scenarios where new heavy degrees of freedom cannot be directly produced at collider energies, their low-energy effects can be systematically described within the Effective Field Theory (EFT) framework in a model-independent manner~\cite{Buchmuller:1985jz,Hagiwara:1993ck}. In this approach, possible deviations from the Standard Model (SM) are parametrized by higher-dimensional operators constructed from SM fields and respecting the underlying $SU(2)_L \times U(1)_Y$ gauge symmetry. 

Since genuine anomalous quartic gauge couplings (aQGCs) can arise without inducing unwanted corrections to triple gauge couplings only at dimension-8 level, the effective Lagrangian relevant for quartic gauge interactions can be expressed as~\cite{Eboli:2006wa,Degrande:2013rea,Degrande:2013yda,Perez:2018kav,Almeida:2020ylr,Durieux:2024zrg}
\begin{eqnarray}\label{lag}
\mathcal{L}_{eff}=\mathcal{L}_{SM}+\sum_{j=0}^{1}\frac{f_{S,j}}{\Lambda^4}\mathcal{O}_{S,j}
+\sum_{j=0}^{7}\frac{f_{M,j}}{\Lambda^4}\mathcal{O}_{M,j}
+\sum_{j=0\atop j\neq 3}^{9}\frac{f_{T,j}}{\Lambda^4}\mathcal{O}_{T,j},
\end{eqnarray}
where $\Lambda$ denotes the characteristic scale of new physics, while $f_{S,j}$, $f_{M,j}$ and $f_{T,j}$ are the corresponding Wilson coefficients. The operator classes $\mathcal{O}_{S,j}$ involve only covariant derivatives of the Higgs doublet, $\mathcal{O}_{M,j}$ include two field strength tensors together with two covariant derivatives of the Higgs field, and $\mathcal{O}_{T,j}$ are constructed solely from electroweak field strength tensors.

In this study, we investigate the neutral vector boson scattering processes
\begin{equation}
\proaa
\quad \text{and} \quad
\proza
\end{equation}
where the $Z$ boson is considered to decay invisibly via $Z\to\nu\bar{\nu}$. 
At high center-of-mass energies, these reactions proceed predominantly through 
$t$-channel electroweak boson exchange, corresponding to vector boson scattering (VBS) topologies.

The tree-level Feynman diagrams of the signal  $\proaa$(top row) and 
$\proza$ (bottom row) processes including anomalous quartic gauge coupling (aQGC) contributions are presented in Fig.~\ref{fd}. Diagrams labeled (a) and (b) correspond to VBS-type topologies, while diagrams (c) and (d) also contain triboson production mechanisms arising from $s$-channel electroweak exchanges. 
The red dots indicate the presence of effective quartic gauge vertices induced by dimension-8 operators. 
For the $\proaa$ process, these anomalous interactions involve 
$\gamma\gamma\gamma\gamma$, $ZZ\gamma\gamma$, and $Z\gamma\gamma\gamma$ vertices. 
Similarly, for the $\proza$ process, the relevant quartic structures include 
$ZZZ\gamma$, $ZZ\gamma\gamma$, and $Z\gamma\gamma\gamma$ interactions. 

The presence of these effective quartic vertices directly reflects the contributions of higher-dimensional operators in the EFT expansion, which modify the neutral gauge boson self-interactions without altering triple gauge couplings at leading order.

Among the dimension-8 operator classes introduced in Eq.~(\ref{lag}), the $\mathcal{O}_{T,j}$ operators directly modify the pure gauge sector and generate anomalous neutral quartic vertices relevant for the final states considered in this work. These operators contribute to the scattering amplitudes via modified gauge boson self-interactions appearing in the subprocesses $\gamma\gamma \to \gamma\gamma$ and $\gamma\gamma \to Z\gamma$, thereby affecting both signal kinematics and total production rates. The explicit form of the $\mathcal{O}_{T,j}$ operators and the corresponding anomalous quartic gauge boson interactions considered in this analysis are summarized in Table~\ref{operators}.

\begin{table}[h!]
\caption{Quartic gauge boson vertices modified by the explicit form of related dimension-8 operators \label{operators} }
\begin{ruledtabular}
\scriptsize{\begin{tabular}{lccccccccc}
Operators&\multicolumn{9}{c}{Vertices} \\
\hline
\multirow{1}{*}{$\mathcal{O}_{T0}=\textrm{Tr}[\widehat{W}_{\mu\nu}\widehat{W}^{\mu\nu}]\times \textrm{Tr}[\widehat{W}_{\alpha\beta}\widehat{W}^{\alpha\beta}]$}\\
\multirow{1}{*}{$\mathcal{O}_{T1}=\textrm{Tr}[\widehat{W}_{\alpha\nu}\widehat{W}^{\mu\beta}]\times \textrm{Tr}[\widehat{W}_{\mu\beta}\widehat{W}^{\alpha\nu}]$}
 & $WWZZ$ & $ZZZZ$ & $WW \gamma Z$ & $WW \gamma \gamma$ & $ZZZ \gamma$ & $ZZ \gamma \gamma$ & $Z\gamma\gamma\gamma$ & $\gamma\gamma\gamma\gamma$ & \\
\multirow{1}{*}{$\mathcal{O}_{T2}=\textrm{Tr}[\widehat{W}_{\alpha\mu}\widehat{W}^{\mu\beta}]\times \textrm{Tr}[\widehat{W}_{\beta\nu}\widehat{W}^{\nu\alpha}]$}\\
\multirow{1}{*}{$\mathcal{O}_{T3}=\textrm{Tr}[\widehat{W}_{\mu\nu}\widehat{W}_{\alpha\beta}]\times \textrm{Tr}[\widehat{W}^{\alpha\nu}\widehat{W}^{\mu\beta}]$} \\
\hline
\multirow{1}{*}{$\mathcal{O}_{T4}=\textrm{Tr}[\widehat{W}_{\mu\nu}\widehat{W}_{\alpha\beta}]\times \widehat{B}^{\alpha\nu}\widehat{B}^{\mu\beta}$}\\
\multirow{1}{*}{$\mathcal{O}_{T5}=\textrm{Tr}[\widehat{W}_{\mu\nu}\widehat{W}^{\mu\nu}]\times \widehat{B}_{\alpha\beta}B^{\alpha\beta}$}
& $ZZZZ$ & $WW \gamma Z$ & $WW \gamma \gamma$ & $ZZZ \gamma$ & $ZZ \gamma \gamma$ & $Z\gamma\gamma\gamma$ & $\gamma\gamma\gamma\gamma$ & \\
\multirow{1}{*}{$\mathcal{O}_{T6}=\textrm{Tr}[\widehat{W}_{\alpha\nu}\widehat{W}^{\mu\beta}]\times \widehat{B}_{\mu\beta}\widehat{B}^{\alpha\nu}$}\\
\multirow{1}{*}{$\mathcal{O}_{T7}=\textrm{Tr}[\widehat{W}_{\alpha\mu}\widehat{W}^{\mu\beta}]\times \widehat{B}_{\beta\nu}\widehat{B}^{\nu\alpha}$}\\
\hline
\multirow{1}{*}{$\mathcal{O}_{T8}=[\widehat{B}_{\mu\nu}\widehat{B}^{\mu\nu}\widehat{B}_{\alpha\beta}\widehat{B}^{\alpha\beta}]$}\\
\multirow{1}{*}{$\mathcal{O}_{T9}=[\widehat{B}_{\alpha\mu}\widehat{B}^{\mu\beta}\widehat{B}_{\beta\nu}\widehat{B}^{\nu\alpha}]$} & $ZZZZ$  & $ZZZ \gamma$ & $ZZ \gamma \gamma$ & $Z\gamma\gamma\gamma$ & $\gamma\gamma\gamma\gamma$ & 
\end{tabular}}
\end{ruledtabular}
\end{table}

At high center-of-mass energies, the scattering amplitudes induced by dimension-8 operators exhibit a strong energy dependence that may lead to a violation of perturbative unitarity if the Wilson coefficients are not sufficiently suppressed. In order to ensure the theoretical consistency of the obtained EFT limits, unitarity constraints on the anomalous quartic gauge couplings have been carefully taken into account following the methodology discussed in Ref.~\cite{Almeida:2020ylr}. In this context, the partial wave unitarity condition is imposed to determine the maximum energy scale up to which the effective field theory description remains valid.

The analytical expressions reported in Ref.~\cite{Almeida:2020ylr} are derived by imposing perturbative unitarity on the $2 \to 2$ vector boson scattering subprocesses induced by the dimension-8 operators. In particular, the scattering amplitudes for subprocesses such as $\gamma\gamma \to \gamma\gamma$ and $\gamma\gamma \to Z\gamma$ are computed including the leading energy contributions arising from the $\mathcal{O}_{T,i}$ operators. Owing to the higher-dimensional nature of these operators, the corresponding cross sections typically grow with energy as $\sigma_{\mathrm{aQGC}}\sim s^{2}/\Lambda^{8}$, potentially leading to a violation of perturbative unitarity at sufficiently high energies.

To quantify this behaviour, the helicity amplitudes are projected onto the partial wave basis according to
\begin{equation}
a_J(s)=\frac{1}{32\pi}\int_{-1}^{1} \mathcal{M}(s,\cos\theta)\,P_J(\cos\theta)\,d(\cos\theta),
\end{equation}
where $P_J(\cos\theta)$ denotes the Legendre polynomials and $\mathcal{M}(s,\cos\theta)$ is the corresponding scattering amplitude. The most stringent constraint is typically obtained from the $J=0$ partial wave component. The perturbative unitarity requirement
\begin{equation}
|{\rm Re}(a_0)| \leq \frac{1}{2}
\end{equation}
is then imposed to determine the maximum center-of-mass energy scale at which the EFT description remains valid for a given value of \couplings.

\begin{figure}[h!]
    \includegraphics[scale=0.7]{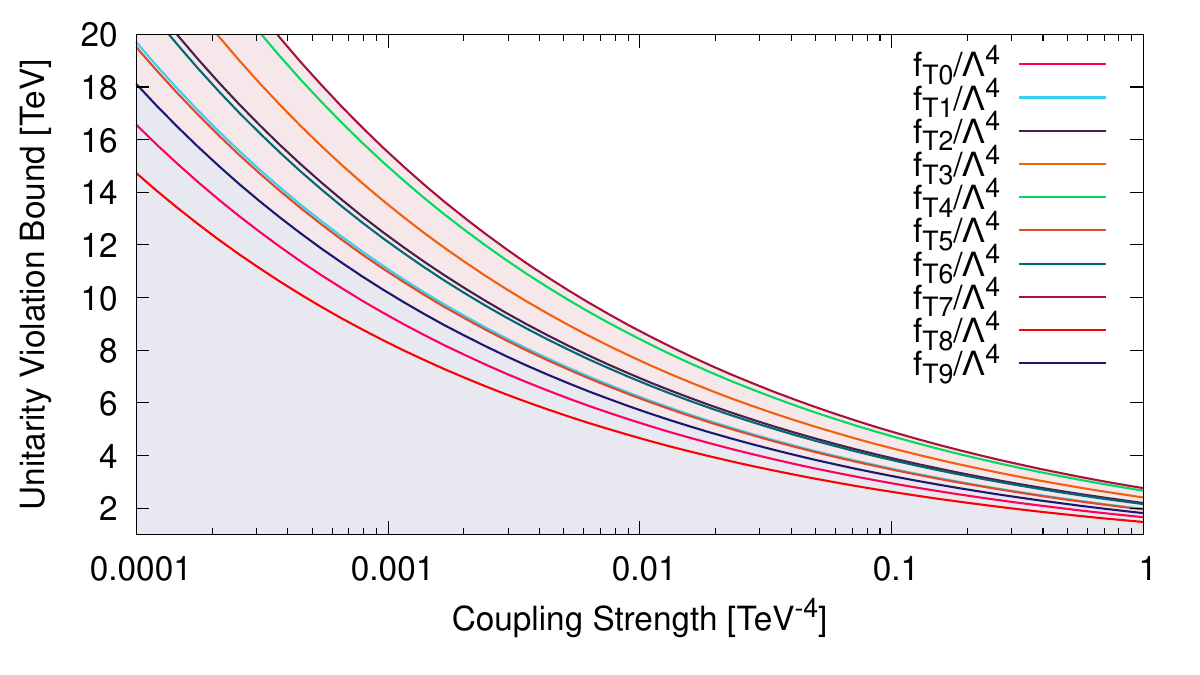}
    \caption{The dependence of the unitarity violation scale on the anomalous coupling strengths \couplings, obtained using the analytical expressions given in Ref.~\cite{Almeida:2020ylr}.}

    \label{fig:unitarity}
\end{figure}

Using the resulting analytical expressions, the unitarity bounds corresponding to each anomalous coupling \couplings are derived. The dependence of the unitarity violation scale on the coupling strength is illustrated in Fig.~\ref{fig:unitarity}. As seen from the figure, increasing values of the aQGC parameters lead to a rapid decrease in the maximum allowed energy scale, indicating the onset of non-perturbative behaviour beyond a certain threshold. These coupling-dependent unitarity limits are subsequently employed in the following sections to restrict the physically meaningful parameter space and to ensure that all reported sensitivity projections are obtained within the regime of EFT validity.

\begin{table}[httb!]
\tiny
\caption{The best observed one dimensional 95\% confidence level limits on dimension-8 anomalous quartic gauge couplings \couplings, obtained by the ATLAS and CMS collaborations in various production channels, with unitarity is preserved. }
\begin{ruledtabular}
    \begin{tabular}{l|ccc|ccc}
        \textbf{Collaboration} & 
        \multicolumn{3}{c}{\textbf{ATLAS ($\sqrt{s} = 13$ TeV)}} & 
         \multicolumn{3}{c}{\textbf{CMS ($\sqrt{s} = 13$ TeV)}} \\
        \hline
        
        Luminosity & 139 $fb^{-1}$ & 140 $fb^{-1}$ & 140 $fb^{-1}$ &138 $fb^{-1}$ & 138 $fb^{-1}$ & 137 $fb^{-1}$ \\
        Channel & $Z(\nu\bar{\nu})\gamma jj$ \cite{ATLAS:2022nru} & $VVjj$ \cite{ATLAS:2025omi} &  VBS+tri-boson\cite{ATLAS:2026wew} & $WVjj$  \cite{CMS:2025dbm} & $W\gamma jj$ \cite{CMS:2022yrl}& $Z\gamma jj$ \cite{CMS:2021gme} \\ 
        & & & & & & \\
        
        \hline
        $f_{T0}/\Lambda^{4}$ (TeV$^{-4}$)  & - & - & [-3.2 ; 2.2]$\times10^{-1}$& [-9.21 ; 7.85]$\times10^{-2}$ & - & -\\
        $f_{T1}/\Lambda^{4}$ (TeV$^{-4}$) & - & - & [-1.6 ; 1.6]$\times10^{-1}$&[-8.63 ; 9.43]$\times10^{-2}$ & - & - \\
        $f_{T2}/\Lambda^{4}$ (TeV$^{-4}$)& -& - & [-6.1 ; 4.2]$\times10^{-1}$&[-2.10 ; 2.14]$\times10^{-1}$ & - & - \\
        $f_{T3}/\Lambda^{4}$ (TeV$^{-4}$)& - & [-8.8 ; 8.8]$\times10^{-1}$ &  -&[-1.91 ; 2.14]$\times10^{-1}$ & - & -\\
        $f_{T4}/\Lambda^{4}$ (TeV$^{-4}$)& -& [-3.03 ; 2.60]$\times10^{0}$&  -&[-8.95 ; 8.28]$\times10^{-1}$ & - & -\\
        $f_{T5}/\Lambda^{4}$ (TeV$^{-4}$)& [-3.4 ; 4.2]$\times10^{-1}$ & -&  -&[-2.65 ; 2.37]$\times10^{-1}$ & - & -\\
        $f_{T6}/\Lambda^{4}$ (TeV$^{-4}$)& - & -  & [-5.5 ; 5.7]$\times10^{-1}$& - & [-2.5 ; 2.7]$\times10^{-1}$ & - \\ 
        $f_{T7}/\Lambda^{4}$ (TeV$^{-4}$)&- & - &  [-5.2 ; 5.6]$\times10^{-1}$&- & [-6.7 ; 7.3]$\times10^{-1}$ & -\\
        $f_{T8}/\Lambda^{4}$ (TeV$^{-4}$)& - & - &  [-2.8 ; 2.8]$\times10^{-1}$&- & - & [-4.7 ; 4.7]$\times10^{-1}$\\ 
        $f_{T9}/\Lambda^{4}$ (TeV$^{-4}$)& - & - & [-5.3 ; 5.3]$\times10^{-1}$& -& - & [-9.1 ; 9.1]$\times10^{-1}$\\
    \end{tabular}
    \label{lhc_limits}
\end{ruledtabular}
\end{table}

The current experimental constraints on the dimension–8 anomalous quartic gauge couplings have been extensively investigated by both the ATLAS and CMS collaborations through various production mechanism at $\sqrt{s}=13$ TeV \cite{ATLAS:2025omi,ATLAS:2022nru, ATLAS:2026wew, CMS:2026job,CMS:2025dbm,CMS:2022yrl, CMS:2021gme,CMS:2021jji,CMS:2020ypo,CMS:2020gfh}.  The most stringent observed one–dimensional 95\% confidence level limits on the \couplings operators obtained from different production channels are summarized in Table~\ref{lhc_limits}, where perturbative unitarity is explicitly preserved in the extraction of the EFT bounds. In particular, the $Z(\nu\bar{\nu})\gamma jj$ electroweak production channel provides direct sensitivity to purely neutral quartic gauge vertices such as $\gamma\gamma\gamma\gamma$ and $ZZ\gamma\gamma$, allowing the extraction of competitive limits on the $f_{T5}$,$f_{T8}$ and $f_{T9}$ operators as reported in Refs.~\cite{ATLAS:2022nru,CMS:2021gme}.
Similarly, semileptonic and fully leptonic VBS topologies such as $WVjj$ and $VVjj$ probe charged–current interactions involving $WW\gamma\gamma$ and $WWZ\gamma$ vertices, leading to enhanced sensitivity to the $f_{T0}$–$f_{T5}$ operator set, as demonstrated in Refs.~\cite{ATLAS:2025omi,CMS:2025dbm}. The combined measurements of vector-boson scattering in numerous final states as well as a tri-boson measurement are used to constrain anomalous electroweak boson quartic self-couplings that result from dimension-8 operators in Ref.~\cite{ATLAS:2026wew}. Furthermore, triboson–like final states such as $W\gamma jj$ offer an experimentally clean environment with reduced QCD backgrounds, enabling CMS to place meaningful constraints on $f_{T6}$ and $f_{T7}$ through the study of electroweak $W\gamma$ production in association with two forward jets~\cite{CMS:2022yrl}.

It is worth emphasizing that the variation in the obtained bounds across different channels originates from the distinct Lorentz structures of the $\mathcal{O}_{T,i}$ operators and their corresponding helicity amplitudes, which modify the VBS kinematic distributions differently depending on the polarization states of the scattered vector bosons. Consequently, each experimental topology exhibits enhanced sensitivity to specific subsets of anomalous couplings, thereby motivating the complementary use of multiple VBS processes in constraining the EFT parameter space.

\section{Cross sections of the processes $\proaa$ and $\proza$}
We investigate the effects of dimension-8 anomalous quartic gauge couplings, parameterized by the coefficients \couplings, through the processes $\proaa$ and $\proza$ at muon colliders. In this study, we restrict our analysis to the $O_{T,i}$ class of operators, which generate genuine anomalous quartic gauge interactions without suppression from electroweak symmetry breaking effects. In contrast, the $O_{M,i}$ operators contribute to these interactions only through terms proportional to $v^2/\Lambda^4$, leading to a reduced impact at high energies. Consequently, their effects are expected to be less significant in the kinematic regime where $\sqrt{s} \gg v$, as considered in this work. The presence of such anomalous interactions, characterized by Lorentz structures different from those of the Standard Model, can modify the kinematic properties of the signal processes and result in observable deviations in the total cross sections.
\begin{figure}[h!]
\includegraphics[scale=0.40]{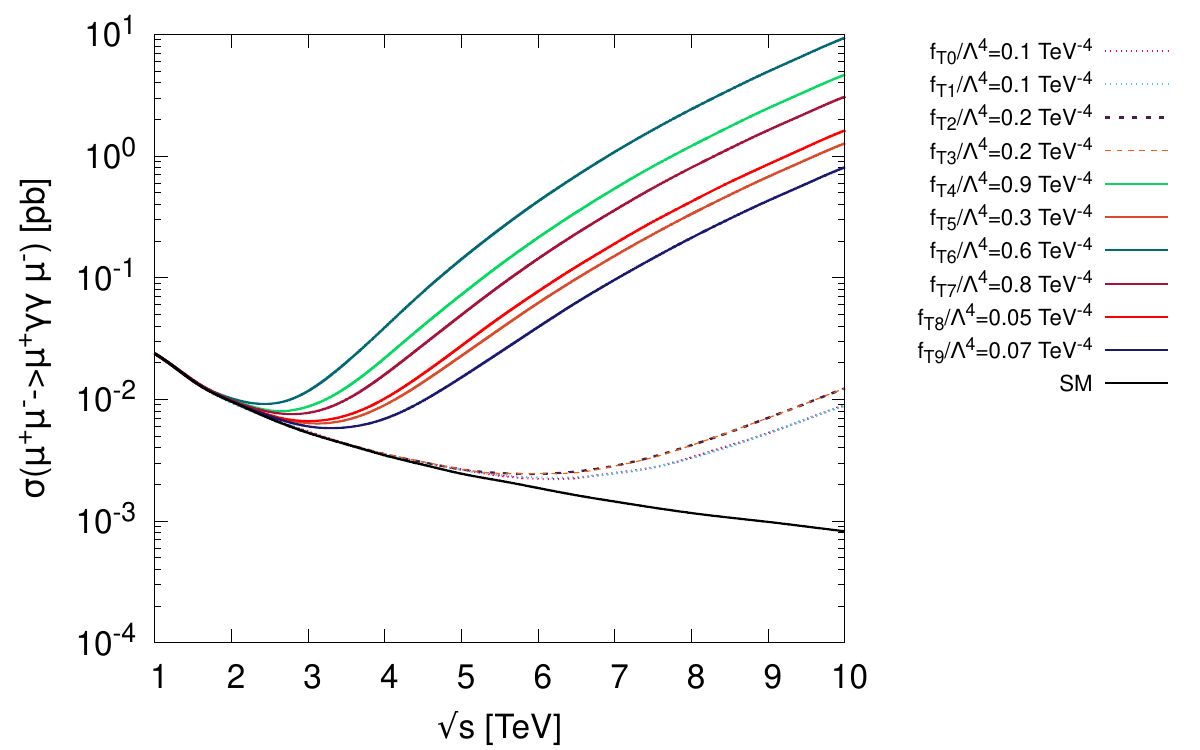}
\includegraphics[scale=0.40]{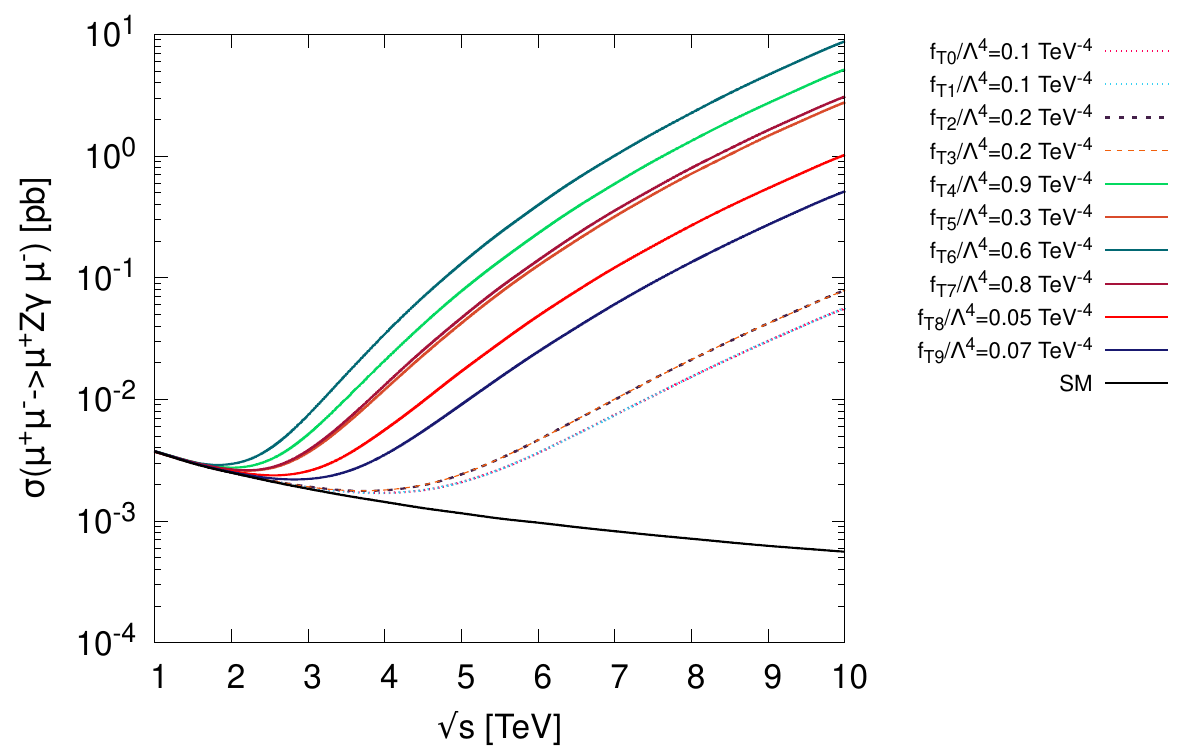}
\caption{The total cross sections of the processes $\proaa$ and $\proza$ as a function of center of mass energy for different anomalous coupling values.}
\label{fig:ecm}
\end{figure}

\begin{figure}[httb!]
\includegraphics[scale=0.40]{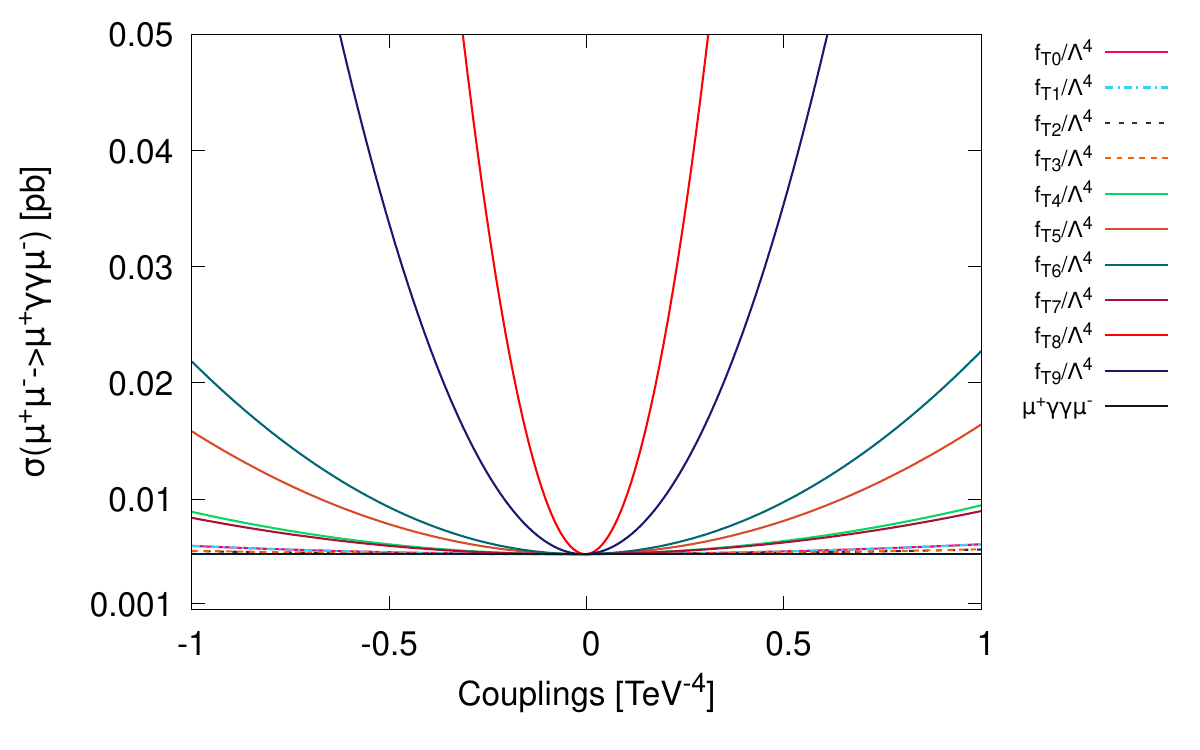}
\includegraphics[scale=0.40]{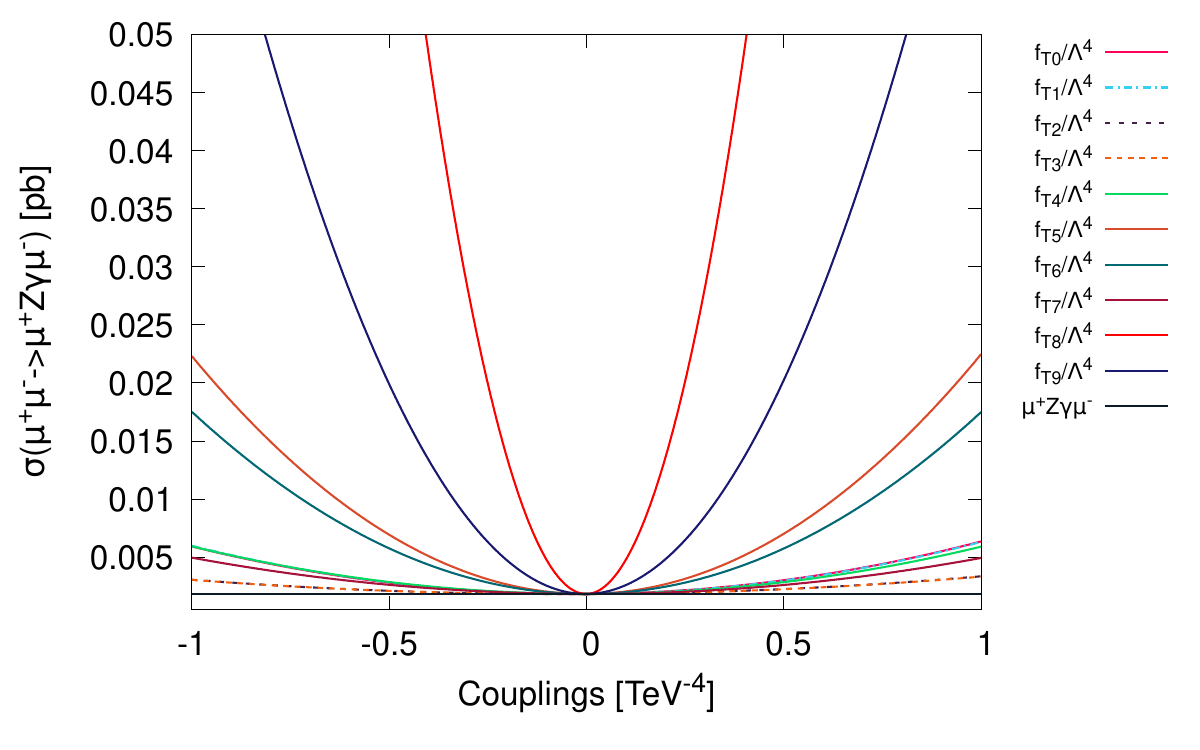}
\includegraphics[scale=0.40]{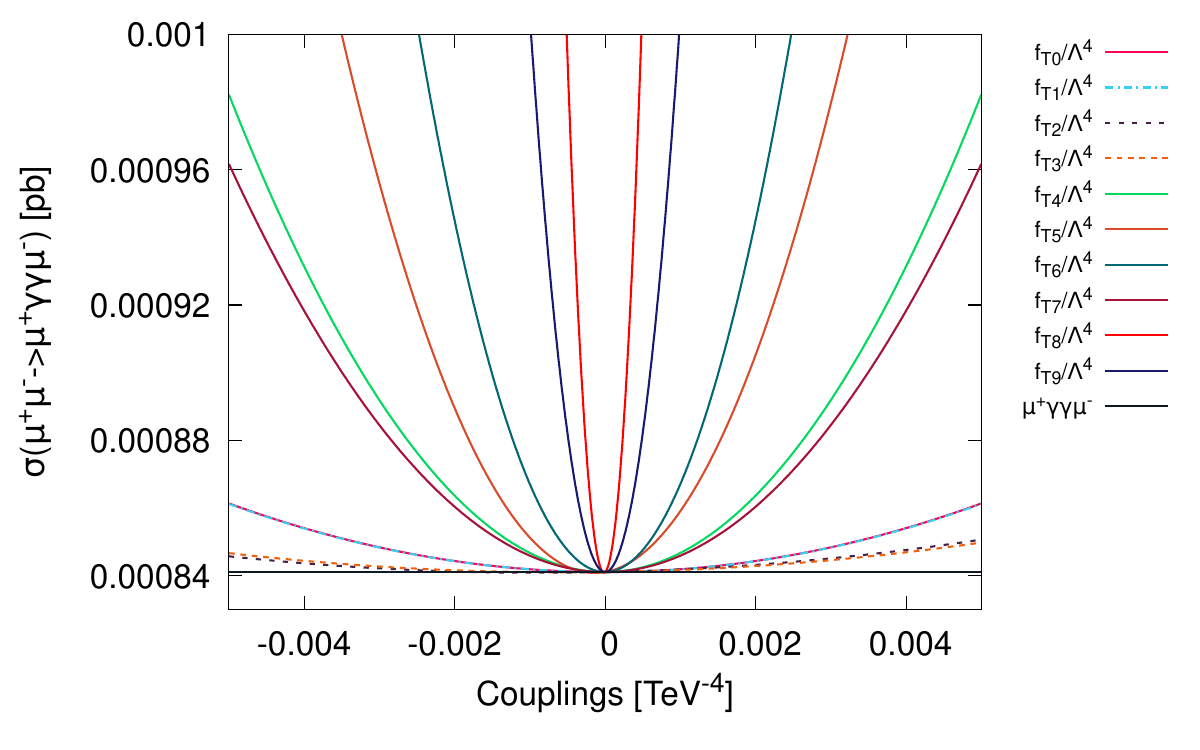}
\includegraphics[scale=0.40]{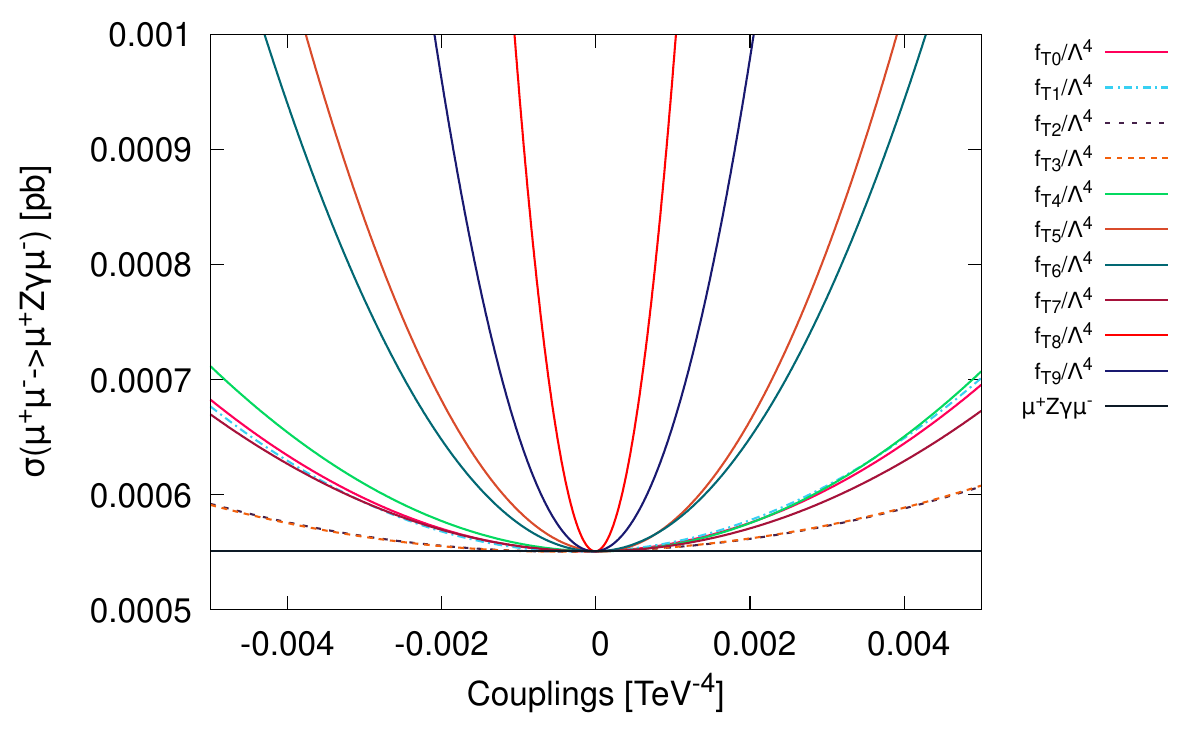}
\caption{The total cross sections as a function of anomalous quartic gauge couplings \couplings for the $\proaa$ and $\proza$ processes at $\com =$ 3 TeV (top row) and $\com$ 10 TeV (bottom row) Muon Collider.}
\label{fig:total_xs}
\end{figure}

Signal and Standard Model (SM) background events with identical final states are generated by varying one anomalous coupling at a time, while fixing all remaining couplings to zero, within a generator-level phase space using \madgraph \texttt{v3.5.6} \cite{Alwall:2014hca}, where the last part of the Effective Lagrangian given in Eq. (\ref{lag}) is implemented through a \ufo \cite{Degrande:2011ua} module generated by \feynrules \cite{Alloul:2013bka} package. This fiducial phase space is established by imposing kinematic cuts on the  final state particles to ensure infrared-safe event generation and compatibility with detector acceptance. Specifically, we apply cuts on the final state particles; transverse momenta $p^{\gamma_i,\mu_i^\pm}_{T} > 10$ GeV, pseudo-rapidity $ |\eta^{\gamma_{i},\mu_i^\pm}| \leq 2.5$, and the minimum distance between each photon and muon ($\Delta R(\gamma_i, \mu^\pm_i) > 0.4$, $\Delta R(\gamma_i, \gamma_j) > 0.4$ and $\Delta R(\mu^+, \mu^-) > 0.4$ ) where $\Delta R(i,j)=[(\Delta\Phi_{i,j})^2 +(\Delta\eta_{i,j})^2]^{1/2}$ with $\Delta\Phi_{i,j}$ and $\Delta\eta_{i,j}$ are the azimuthal angle and the pseudo-rapidity difference between any two final state particles.

The center-of-mass energy dependence of the total cross sections for the processes $\proaa$ and $\proza$ is illustrated in Fig.~\ref{fig:ecm} for representative values of the anomalous quartic gauge couplings, chosen in accordance with the current best experimental limits reported by the ATLAS and CMS collaborations. As expected from the higher–dimensional nature of the $\mathcal{O}_{T,i}$ operators, the corresponding EFT contributions exhibit a pronounced growth with increasing center-of-mass energy, scaling approximately as $s^{2}/\Lambda^{8}$.
While the Standard Model tends to decrease with energy as a consequence of electroweak $t$-channel exchange, the presence of anomalous quartic gauge interactions leads to a significant enhancement of the total cross section at higher energies. In particular, it is observed that deviations from the Standard Model expectation begin to emerge already at around $\sqrt{s}=3$ TeV, where the anomalous contributions become comparable to the SM background for representative benchmark coupling values

This energy scale therefore represents the onset of measurable sensitivity to dimension–8 effects in muon–initiated vector boson scattering processes.
On the other hand, at $\sqrt{s}=10$ TeV the cross section receives increasingly significant contributions from EFT-induced interactions for several operator choices, yielding deviations of more than one order of magnitude with respect to the Standard Model prediction. This substantial enhancement significantly improves the discovery potential for anomalous quartic gauge couplings and allows for the exploration of parameter regions well beyond the reach of the current LHC measurements. Consequently, the benchmark center-of-mass energies of 3 TeV and 10 TeV are adopted throughout this work as representative operating points for future muon collider scenarios, corresponding respectively to the threshold of EFT sensitivity and to the regime of maximal anomalous coupling enhancement within the perturbative unitarity bounds. These observations directly motivate the choice of the collider energies considered in the following analysis.

In order to further quantify the sensitivity of the considered processes to anomalous quartic gauge interactions, the dependence of the total cross sections on the dimension–8 operator coefficients \couplings is presented in Fig.~\ref{fig:total_xs} for both $\proaa$ and $\proza$ productions at $\sqrt{s}=3$ TeV and $\sqrt{s}=10$ TeV. As clearly observed, the cross sections exhibit a characteristic polynomial dependence on the anomalous couplings, arising from both the interference between the SM and EFT amplitudes (linear terms) and the pure dimension-8 operator contributions (quadratic terms). At $\sqrt{s}=3$ TeV, deviations from the Standard Model prediction remain relatively moderate within the experimentally allowed parameter ranges, indicating that this energy scale corresponds to the initial sensitivity regime where EFT effects begin to manifest themselves. Nevertheless, certain operators, particularly those associated with field–strength tensor structures, already induce visible enhancements in the total production rates even for small coupling values.
In contrast, at $\sqrt{s}=10$ TeV the impact of anomalous quartic gauge couplings becomes significantly more pronounced. The cross sections demonstrate a much steeper dependence on the coupling strengths, reflecting the strong energy growth of the EFT contributions proportional to $s^{2}/\Lambda^{8}$. Consequently, even small deviations from the Standard Model coupling configuration lead to substantial enhancements in the predicted event rates, thereby considerably improving the probing potential of the future muon collider for dimension–8 operators.
These observations consistently support the choice of $\sqrt{s}=3$ TeV and $\sqrt{s}=10$ TeV as representative benchmark energies, corresponding respectively to the threshold of measurable EFT sensitivity and to the regime where anomalous quartic gauge coupling effects can be maximally amplified while remaining within the perturbative unitarity limits.

\section{Monte Carlo Simulations and Multivariate Analysis Methodology}\label{secIII}
In this section, we give details of the analysis to investigate the effects of anomalous quartic gauge couplings \couplings, parametrized by dimension-8 operators, on the neutral anomalous quartic gauge vertices ($ZZ\gamma\gamma$, $Z\gamma\gamma\gamma$, $\gamma\gamma\gamma\gamma$). Signal processes $\proaa$ and $\proza$ along with the relevant background proceses are considered at $\sqrt{s}=3$ TeV with $L_{\mathrm{int}}=1\,\mathrm{ab^{-1}}$ and $\sqrt{s}=10$ TeV with $L_{\mathrm{int}}=10\,\mathrm{ab^{-1}}$ Muon Collider.

The dominant Standard Model background contributions to the $\gamma\gamma\mu\mu$ and $Z\gamma\mu\mu$ final states are summarized in Table~\ref{bpcs}. The corresponding leading–order cross sections are calculated at $\sqrt{s}=3$ TeV and $\sqrt{s}=10$ TeV using the \texttt{MadGraph5\_aMC@NLO} event generator framework.
The considered background processes are primarily selected based on their ability to reproduce the same visible final–state topology as the signal process investigated in this work. In particular, the $\proaa$ signal process involves two isolated photons accompanied by oppositely charged muons in the final state. Consequently, Standard Model diboson production mechanisms with additional photons and leptons, such as $W^{+}W^{-}\gamma\gamma$, $Z\gamma\gamma$, and $W\gamma\gamma$ production, constitute reducible backgrounds that may enter the signal region through additional leptonic decays or mis-reconstruction effects. Similarly, the $\proza$ production, particularly when the $Z$ boson undergoes an invisible decay via $Z\rightarrow\nu\bar{\nu}$, leads to a final–state configuration characterized by a single high–energy photon accompanied by missing transverse energy and charged leptons. In this case, processes such as $ZZ\mu^{+}\mu^{-}$ and $Z\gamma$ production may partially mimic the signal topology through neutrino-induced missing momentum when accompanied by additional leptonic activity. In addition to electroweak diboson production, top–quark pair productions in association with photons (e.g., $t\bar{t}\gamma$ and $t\bar{t}\gamma\gamma$) are also included in the analysis. These processes can generate multi–lepton and through initial- and final-state radiation associated with the leptonic decays of the $W$ bosons originating from top-quark decays, thus leading to non–negligible contamination of the signal region, especially at higher center–of–mass energies.

Overall, these backgrounds represent the dominant irreducible and reducible Standard Model contributions capable of imitating the kinematic characteristics of the $\gamma\gamma\mu\mu$ and $Z\gamma\mu\mu$ final states under consideration.

\begin{table}[h]
\begin{ruledtabular}
\caption{Standard Model contributions to the signal processes and the considered background processes and their corresponding leading-order cross sections at a muon collider with center-of-mass energies $\sqrt{s}=3$ TeV and $\sqrt{s}=10$ TeV. The event samples are generated using \texttt{MadGraph5\_aMC@NLO}. \label{bpcs}}
    \begin{tabular}{lcc}
         Process &
        \begin{tabular}[c]{@{}c@{}}Cross-Section [pb] \\  $\sqrt{s}=3~\text{TeV}$\end{tabular} &
        \begin{tabular}[c]{@{}c@{}}Cross-Section [pb] \\  $ \sqrt{s}=10~\text{TeV}$\end{tabular} \\
       \hline
        $\mu^+\mu^-\to\mu^+\gamma\gamma\mu^-$ ($f_{T,i}/\Lambda^{4} = 0$) & 5.45 $\times 10^{-2}$& 8.44 $\times 10^{-4}$ \\
        $\mu^+\mu^-\to\mu^+Z\gamma\mu^-$ ($f_{T,i}/\Lambda^{4} = 0$) & 1.83  $\times 10^{-3}$ & 5.54  $\times 10^{-4}$\\
        $\mu^+\mu^-\to W^+W^-\gamma\gamma$ & 2.25  $\times 10^{-3}$ & 7.04 $\times 10^{-4}$ \\
        $\mu^+ \mu^- \to W^+ W^-\mu^+\mu^-$ & 2.67  $\times 10^{-2}$ & 1.21  $\times 10^{-2}$\\
        $\mu^+ \mu^- \to ZZ\mu^+\mu^-$ & 1.55 $\times 10^{-4}$& 9.54  $\times 10^{-5}$\\
        $\mu^+ \mu^-\to t\bar{t}\gamma\gamma$ & 3.57  $\times 10^{-4}$& 2.97  $\times 10^{-5}$ \\
        $\mu^+\mu^- \to Z\gamma\gamma$ & 5.76  $\times 10^{-3}$& 7.50  $\times10^{-4}$\\
        $\mu^+\mu^- \to W\gamma\gamma$ &2.79 $\times10^{-2}$ & 5.04$\times10^{-3}$ \\
        $\mu^+\mu^- \to t\bar{t}\gamma$ & 4.67 $\times10^{-2}$& 4.21$\times10^{-3}$ \\
        $\mu^+\mu^- \to Z\gamma$ & 4.00$\times10^{-3}$ & 5.45$\times10^{-4}$ \\
    \end{tabular}
\end{ruledtabular}
\end{table}

To conduct a detailed analysis, approximately $10^6$ events are generated for each signal and background process using {\sc MadGraph5\_aMC@NLO}. This is done for each background process as well as signal, corresponding to different values of the $f_{Ti}/\Lambda^{4}$ couplings. Events are then passed through to {\sc Pythia 8.20} \cite{Sjostrand:2014zea} for parton showering and hadronization. Detector response effects, parametrized in terms of resolution functions and reconstruction efficiencies, are simulated using  {\sc Delphes 3.4.2} \cite{deFavereau:2013fsa} software using the dedicated muon collider detector cards, namely \verb|delphes_card_MuonColliderDet.tcl| for $\sqrt{s}=3$ TeV and \verb|delphes_card_MUSICDet_target.tcl| for $\sqrt{s}=10$ TeV. 
\begin{table}[httb!]
    \caption{Event selection and applied kinematic cuts used in the analysis for the $\proaa$ process at the $\sqrt{s} = 3$ TeV ($\sqrt{s} = 10 $ TeV ) Muon Collider before the multivariate analysis .} 
    \begin{ruledtabular}
    \begin{tabular}{l l }
        \textbf{Cuts} &  Parameters \\
        \hline
        \hline
        Event Pre-selection &  $N_{\mu}> 1$ \&  $N_{\gamma} > 1$  \& $Charge(\mu_{1}) \neq Charge(\mu_{2})$  \\ 
        \hline
        & $p_T^{\gamma_1} > 10$ GeV,\quad $p_T^{\gamma_2} > 10$ GeV  \\
        Kinematic Cuts  & $p_T^{\mu_1} > 20$ GeV,\quad $p_T^{\mu_2} > 10$ GeV    \\
         & \( |\eta^{\gamma_1,\gamma_2,\mu_1,\mu_2}| \leq 2.5\) \\
         & $\Delta R(\mu_{1}, \mu_{2}) > 0.4$ \\
         \hline
        Reconstructed Cuts & $m_{\mu_1 \mu_2} < 2000 (6000)$ GeV \\
        & $\gamma\gamma\text{-cent} < 0.4$ \\
    \end{tabular}
    \end{ruledtabular}
    \label{cut_table_mumuaa}
\end{table}

\begin{table}[httb!]
    \caption{Event selection and applied kinematic cuts used in the analysis for the $\proza$ process at the $\sqrt{s} = 3$ TeV ($\sqrt{s} = 10 $ TeV ) Muon Collider before the multivariate analysis .} 
    \begin{ruledtabular}
    \begin{tabular}{l l }
        \textbf{Cuts} &  Parameters \\
        \hline
        \hline
        Event Pre-selection &  $N_{\mu}> 1$ \&  $N_{\gamma} > 0$  \& $Charge(\mu_{1}) \neq Charge(\mu_{2})$  \\ 
        \hline
        & $p_T^{\gamma_1} > 10$ GeV  \\
      Kinematic Cuts  & $p_T^{\mu_1} > 20$ GeV,\quad $p_T^{\mu_2} > 10$ GeV    \\
         & \( |\eta^{\gamma_1,\mu_1,\mu_2}| \leq 2.5\) \\
         & $\Delta R(\mu_{1}, \mu_{2}) > 0.4$ \\
         \hline
        Reconstructed Cuts & $m_{\mu_1 \mu_2} < 2000 (6000)$ GeV \\
        & $\gamma\text{-cent} < 0.4$ \\
    \end{tabular}
    \end{ruledtabular}
    \label{cut_table_mumuaz}
\end{table}

In order to enhance the sensitivity to anomalous quartic gauge couplings while suppressing the dominant Standard Model background contributions, a set of event pre–selection and kinematic requirements is applied prior to the multivariate analysis for both signal processes under consideration.

For the $\proaa$ production, events are first required to contain at least two isolated photons ($N_{\gamma} > 1$) and two oppositely charged muons ($N_{\mu} > 1$) in the final state. This pre–selection ensures the presence of the visible topology associated with the signal process while significantly reducing contributions from purely leptonic or single–photon background channels. In addition, transverse momentum requirements are imposed on the leading and sub–leading photons ($p^{\gamma_{1},\gamma_{2}}_{T} > 10$ GeV) as well as on the muons ($p^{\mu_{1}}_{T} > 20$ GeV and $p^{\mu_{2}}_{T} > 10$ GeV) in order to suppress non–VBS–like configurations and enhance the selection of events consistent with electroweak vector boson scattering topologies. A pseudorapidity requirement of $|\eta|\leq 2.5$ is also applied to all reconstructed objects to ensure detector acceptance. Furthermore, an angular separation criterion of $\Delta R(\mu_1,\mu_2)>0.4$ is enforced to ensure a well–separated forward muon system characteristic of VBS–induced scattering topologies

Similarly, for the $\proza$ process, events are required to contain at least one reconstructed photon together with two oppositely charged muons satisfying $N_{\mu}>1$ and $N_{\gamma}>0$. The transverse momentum requirements of $p_{T}^{\gamma_1}>10$ GeV, $p^{\mu_{1}}_{T} > 20$ GeV and $p^{\mu_{2}}_{T} > 10$ GeV are imposed together with the pseudorapidity acceptance $|\eta^{\gamma_{1},\mu_{1},\mu_{2}}|\leq2.5$. In order to ensure a well–resolved final state configuration and to suppress collinear backgrounds, a minimum angular separation of $\Delta R(\mu_{1},\mu_{2})>0.4$ is also required.

Following the application of these baseline selections, additional topology–based reconstructed cuts are introduced for both processes. In particular, an upper bound on the invariant mass of the muon pair is imposed to select events with forward-scattered muons originating from t-channel electroweak exchange and to suppress centrally produced dilepton configurations arising from non-scattering background processes.

Moreover, the photon centrality variables, namely $\gamma\gamma$–cent for the $\proaa$ process and $\gamma$–cent for the $\proza$ process, effectively quantify the degree to which the photon or diphoton system is produced centrally within the rapidity gap formed by the outgoing muons, which is a characteristic feature of electroweak vector boson scattering processes. The photon centrality observable is defined as
\[
\gamma\text{-cent} = \left|\frac{y(\gamma_{1}) - \frac{1}{2}\left[y(\mu_{1}) + y(\mu_{2})\right]}{y(\mu_{1})-y(\mu_{2})}\right|,
\]
while for the diphoton system it is given by
\[
\gamma\gamma\text{-cent} = \left|\frac{y(\gamma_{1}\gamma_{2}) - \frac{1}{2}\left[y(\mu_{1}) + y(\mu_{2})\right]}{y(\mu_{1})-y(\mu_{2})}\right|.
\]
where $y(\mu_1)$ and $y(\mu_2)$ denote the rapidities of
the leading and sub–leading muons, respectively; $y(\gamma_1)$ and $y(\gamma_{1},\gamma_{2})$ denote the rapidities of the leading photon and di-photon system.

\begin{figure}[h!]

\includegraphics[scale=0.40]{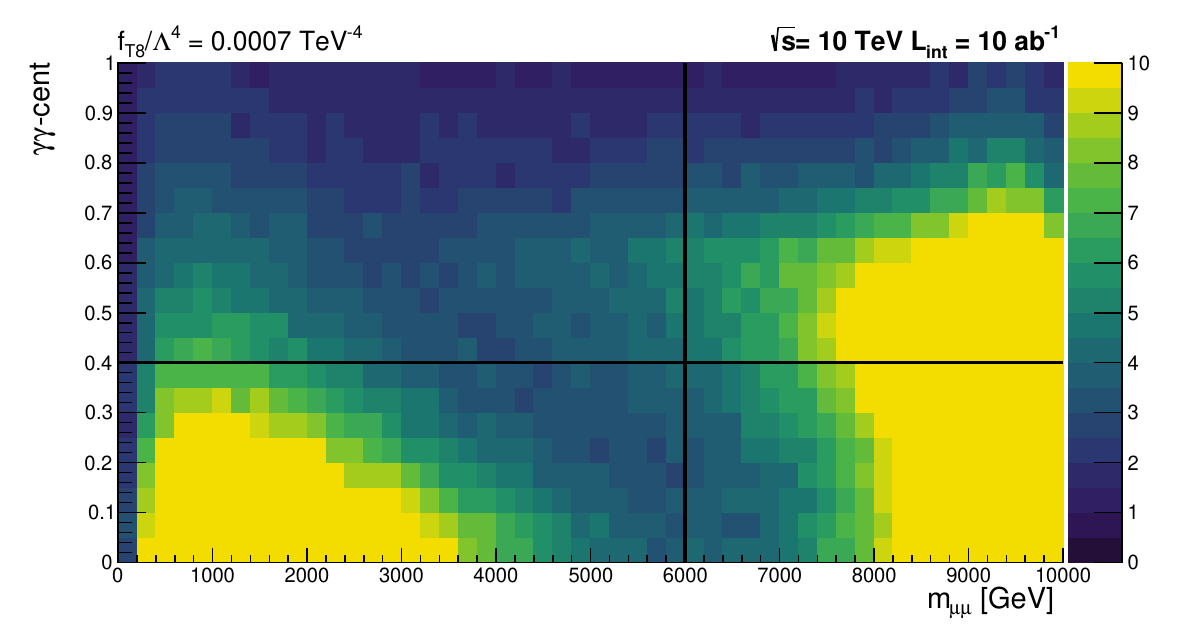}
\includegraphics[scale=0.40]{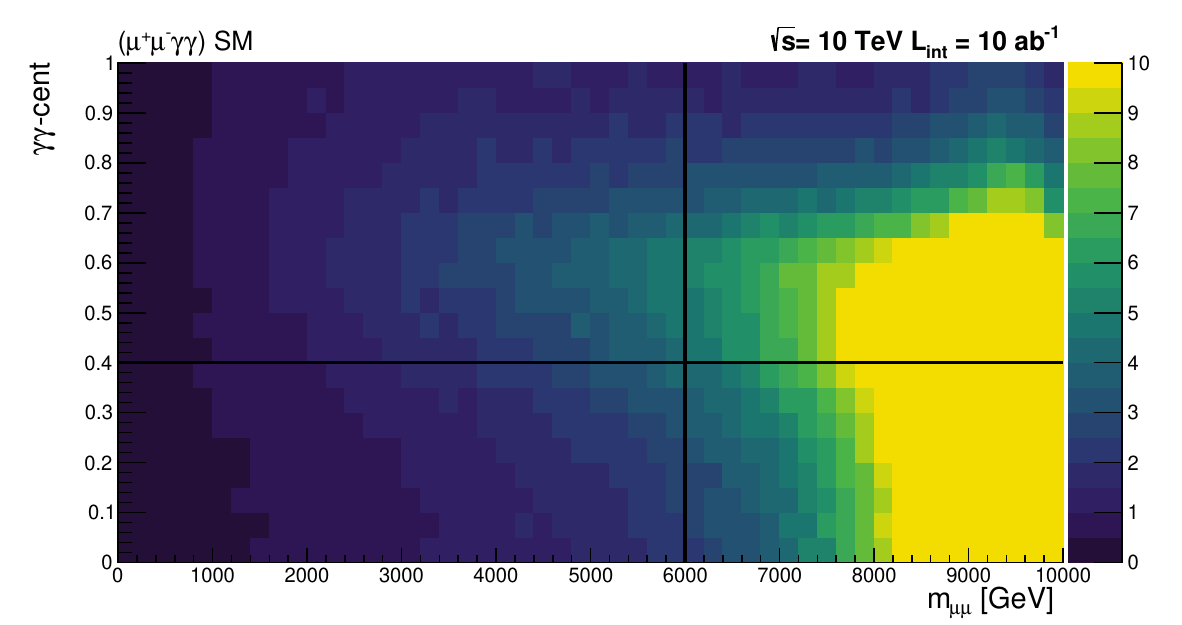}
\includegraphics[scale=0.40]{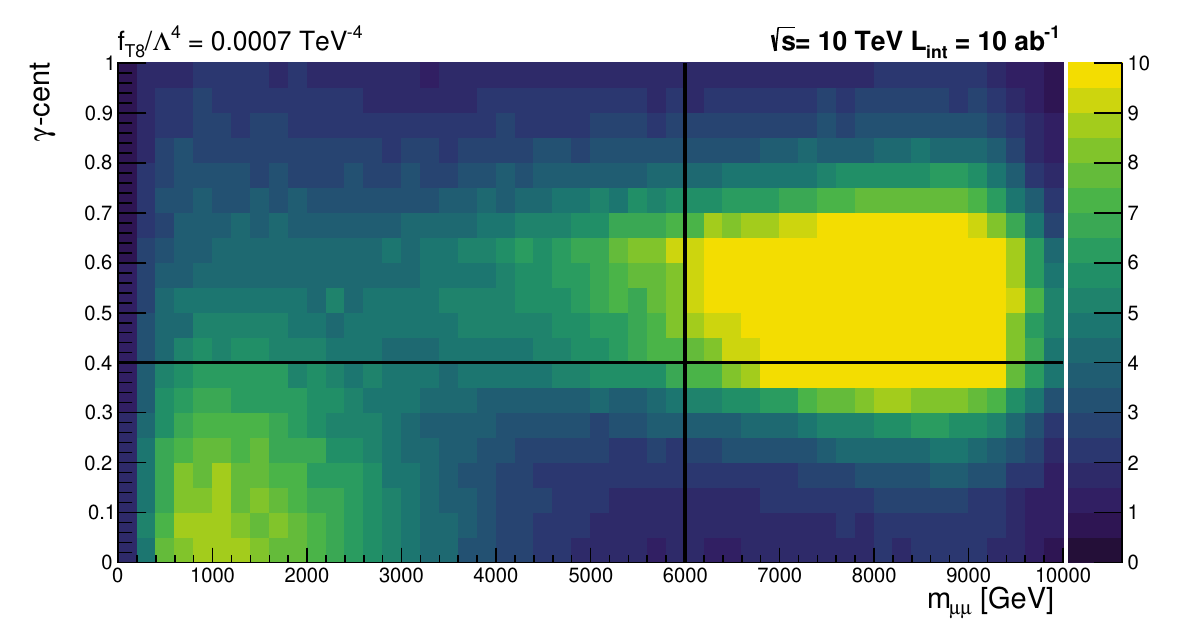}
\includegraphics[scale=0.40]{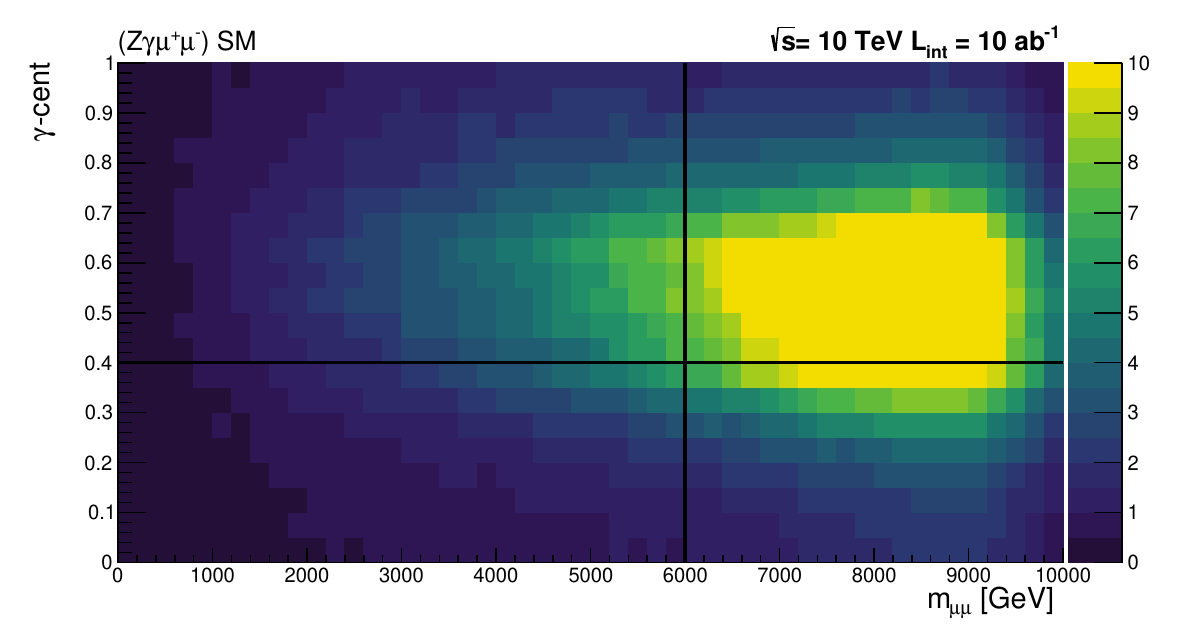}

\caption{Two–dimensional distributions of the photon centrality observables 
($\gamma\gamma$–cent for $\proaa$ and $\gamma$–cent for $\proza$) 
as a function of the dilepton invariant mass $m_{\mu\mu}$ for the signal $f_{T8}/\Lambda^{4} = 0.0007$ and corresponding SM at $\sqrt{s}=10$ TeV Muon Collider. }
\label{muu_cent_aa}
\end{figure}

In Fig.~\ref{muu_cent_aa}, the two–dimensional normalized distributions of the photon centrality observables ($\gamma\gamma$–cent and $\gamma$–cent) as a function of the di–muon invariant mass $m_{\mu\mu}$ are presented for the signal benchmark point $f_{T8}/\Lambda^{4}=0.0007$ TeV$^{-4}$ together with the corresponding SM background processes at $\sqrt{s}=10$ TeV for the $\proaa$ (top row) and $\proza$ (bottom row) processes.

The corresponding two–dimensional distributions reveal a clear correlation between the invariant mass of the muon pair and the spatial localization of the photon(s) in the detector acceptance. 
Anomalous signal contributions are observed to exhibit a distinct population pattern in the ($m_{\mu\mu}$, centrality) phase space compared to the dominant Standard Model backgrounds. Based on this observed behaviour, optimized upper bounds on the dilepton invariant mass are introduced in order to enhance the signal–to–background separation. The selection requirements of $m_{\mu\mu} < 2000~\mathrm{GeV}$ for $\sqrt{s}=3$ TeV and $m_{\mu\mu} < 6000~\mathrm{GeV}$ for $\sqrt{s}=10$ TeV are therefore determined from the joint analysis of the $m_{\mu\mu}$ and centrality distributions. These energy–dependent invariant mass thresholds significantly suppress high–mass dilepton backgrounds while preserving the characteristic event topology of the signal processes.

Furthermore, signal events originating from VBS–like electroweak topologies tend to populate the central detector region more prominently compared to the corresponding Standard Model backgrounds. Accordingly, optimized selection criteria of $\gamma\gamma$–cent $<0.4$ and $\gamma$–cent $<0.4$ are applied in order to maximize the signal–to–background separation.

All generated signal and background events are subsequently required to pass the aforementioned selection criteria before being used in the multivariate analysis. The surviving event samples are then processed within the TMVA framework, where a Boosted Decision Tree (BDT) algorithm is employed to further enhance the discrimination power between the anomalous signal contributions and the remaining Standard Model backgrounds.



\begin{figure}[h!]
\includegraphics[scale=0.40]{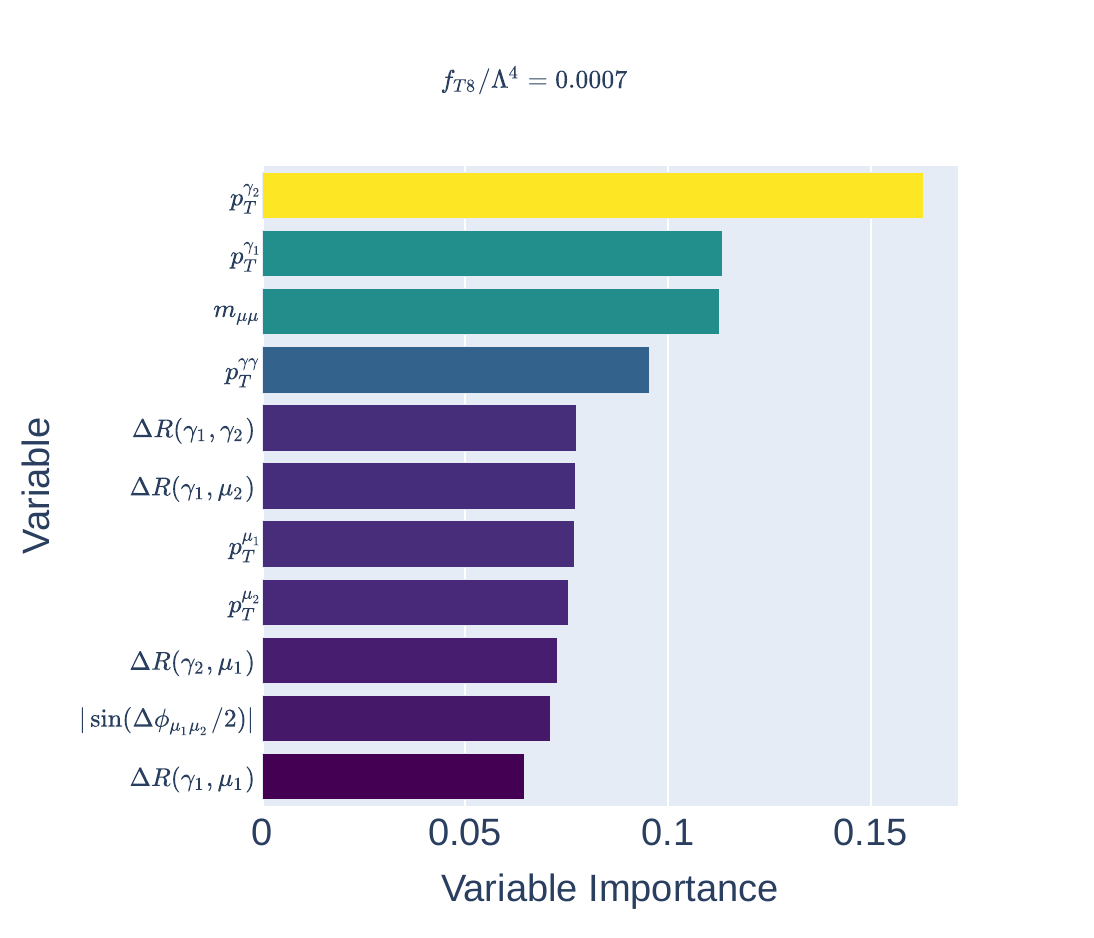}
\includegraphics[scale=0.40]{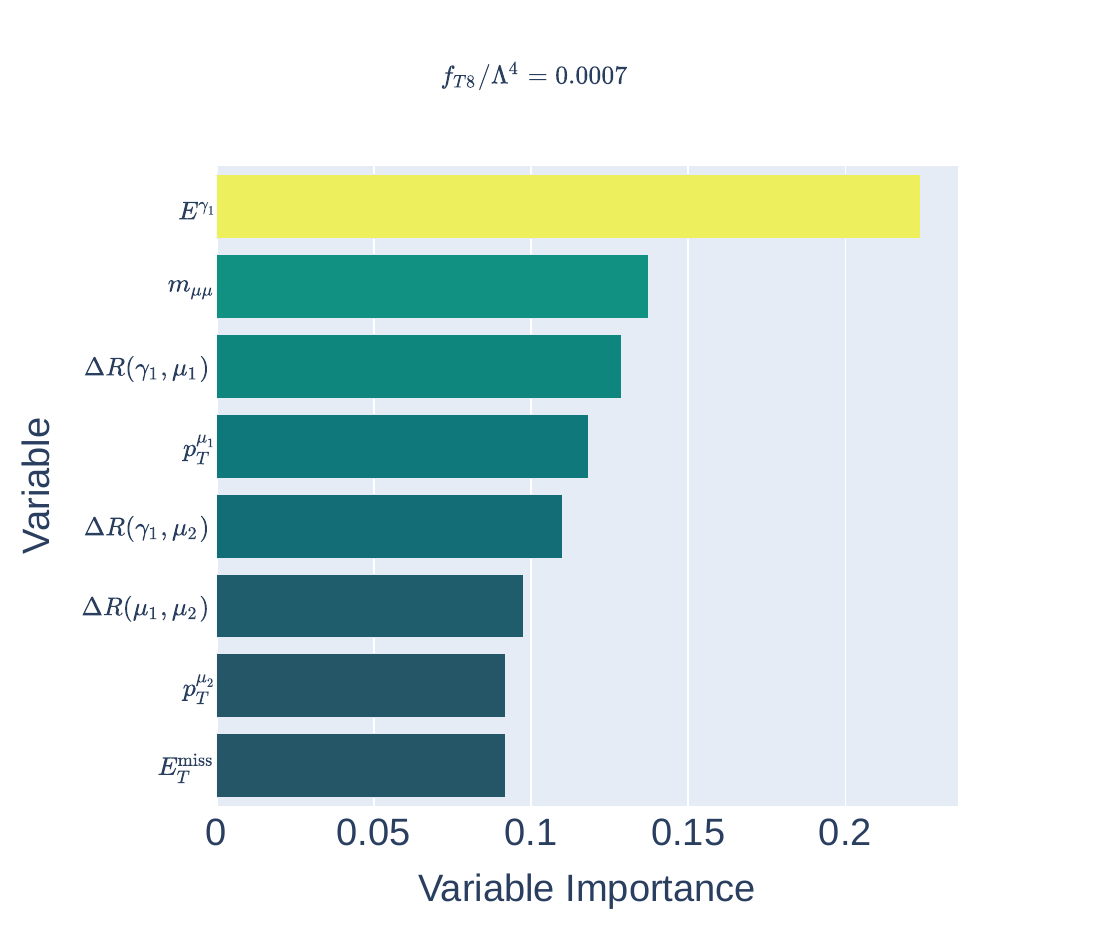}

\caption{The variable list for the processes $\proaa$ (on the left) and $\proza$ (on the right) used in the analysis and their relative importances at $\com$ = 10 TeV Muon Collider.}
\label{var_imp}
\end{figure}

In order to further enhance the sensitivity to signal processes especially with small coupling values and to distinguish them from the remaining background processes after the applied cuts given in Table \ref{cut_table_mumuaa}, are utilized to train a Boosted Decision Tree (BDT) algorithm for each signal process. A BDT is initially trained using a set of object-level kinematical observables of the final-state particles, together with reconstructed kinematic observables, employing the default TMVA hyperparameter configuration. One of the key aspects of BDTs is to identify which input observables contribute most to the classifier’s decision, thereby providing insight into the discriminating power of each observable between signal and background processes. The relative importance of the input observables are extracted from the default BDT training for each anomalous couplings. For each coupling and their values, the most 15 influential observables are identified based on their relative importance. A common subset of input variables is subsequently selected. Only those observables that are shared among all coupling scenarios are retained. Based on this ranking, the input variable set is reduced to the most discriminating observables, which are later used for the final BDT training and hyperparameter optimization. The relative importance of the input observables is extracted from the default BDT training is given in Fig \ref{var_imp} for the processes $\proaa$ and $\proza$ given a benchmark coupling value $f_{T8} = 0.0007$ TeV$^{-4}$ at $\sqrt{s} = 10$ TeV Muon Collider. The ranking reveals that the transverse momentum of the final state particles and angular relations between them provide the strongest discrimination between the anomalous signal contributions and the remaining Standard Model background processes.

Following the implementation of the TMVA framework and the training of the Boosted Decision Tree (BDT) classifier, the relative importance of the input observables has been evaluated in order to identify the most discriminating kinematic variables for both signal processes under consideration. The resulting rankings of the ten most influential variables are presented in Fig.~\ref{var_imp} for the benchmark coupling value $f_{T8}/\Lambda^{4}=0.0007~\mathrm{TeV}^{-4}$.

For the $\proaa$ process, the transverse momenta of the leading and sub–leading photons, $p_{T}^{\gamma_1}$ and $p_{T}^{\gamma_2}$, provide the dominant contribution to the separation power between the anomalous signal and the Standard Model background. This behaviour is expected, as the presence of dimension–8 operators lead to an enhancement of high–energy event configurations, resulting in harder photon spectra compared to the Standard Model background.
In addition, the dilepton invariant mass $m_{\mu\mu}$ and the diphoton transverse momentum $p_{T}^{\gamma\gamma}$ emerge as highly sensitive observables, reflecting the strong correlation between the EFT–induced momentum transfer and the overall event topology. Angular separation variables such as $\Delta R(\gamma_i,\mu_j)$ and $\Delta R(\gamma_1,\gamma_2)$ further improve the classification performance by exploiting the distinct spatial configurations of the final–state particles in signal events.

A similar trend is observed for the $\proza$ process, where the photon energy $E^{\gamma}$ constitutes the most significant input variable. The dilepton invariant mass $m_{\mu\mu}$ again plays a crucial role in distinguishing the EFT contributions from the Standard Model expectations, in agreement with the previously discussed invariant mass–centrality correlations. Moreover, angular observables such as $\Delta R(\gamma,\mu_1)$ and $\Delta R(\mu_1,\mu_2)$, together with the transverse momenta of the muons and the missing transverse energy $E_{T}^{\mathrm{miss}}$, provide additional sensitivity by capturing the characteristic event topology associated with the neutrino decay of the $Z$ boson.

These results demonstrate that momentum–based observables and angular separation variables collectively offer substantial discrimination capability for probing anomalous quartic gauge interactions. The identified set of input variables is therefore employed in the subsequent multivariate analysis to maximize the sensitivity of the future muon collider to dimension–8 EFT operators.

In the multivariate analysis, the BDT classifier is trained using the most influential kinematical and reconstructed observables constructed from the leading and sub–leading photons ($\gamma_{1}$, $\gamma_{2}$) and muons ($\mu_{1}$, $\mu_{2}$). The training is performed using adaptive boosting (AdaBoost) with 450 decision trees, a maximum tree depth of 3, a minimum of 2.5 events per terminal leaf, 20 cut optimization steps, and a learning rate of 0.5. The full event sample is divided into statistically independent subsets, with 50\% of the data used for training and the remaining 50\% reserved for testing. Background events are combined after being properly weighted according to their respective production cross sections, and all events are required to satisfy the selection criteria summarized in Tables~\ref{cut_table_mumuaa} and \ref{cut_table_mumuaz}.

In order to preserve perturbative unitarity within the EFT framework, the clipping method is employed. This approach suppresses EFT contributions beyond a defined energy scale $\Lambda_{FF}$ determined individually for each anomalous coupling coefficient, as discussed in the previous section. The corresponding clipping thresholds are applied to the invariant mass of the two–photon system and to the transverse momentum of the leading photon in the event selection. 

\begin{figure}
\includegraphics[scale=0.40]{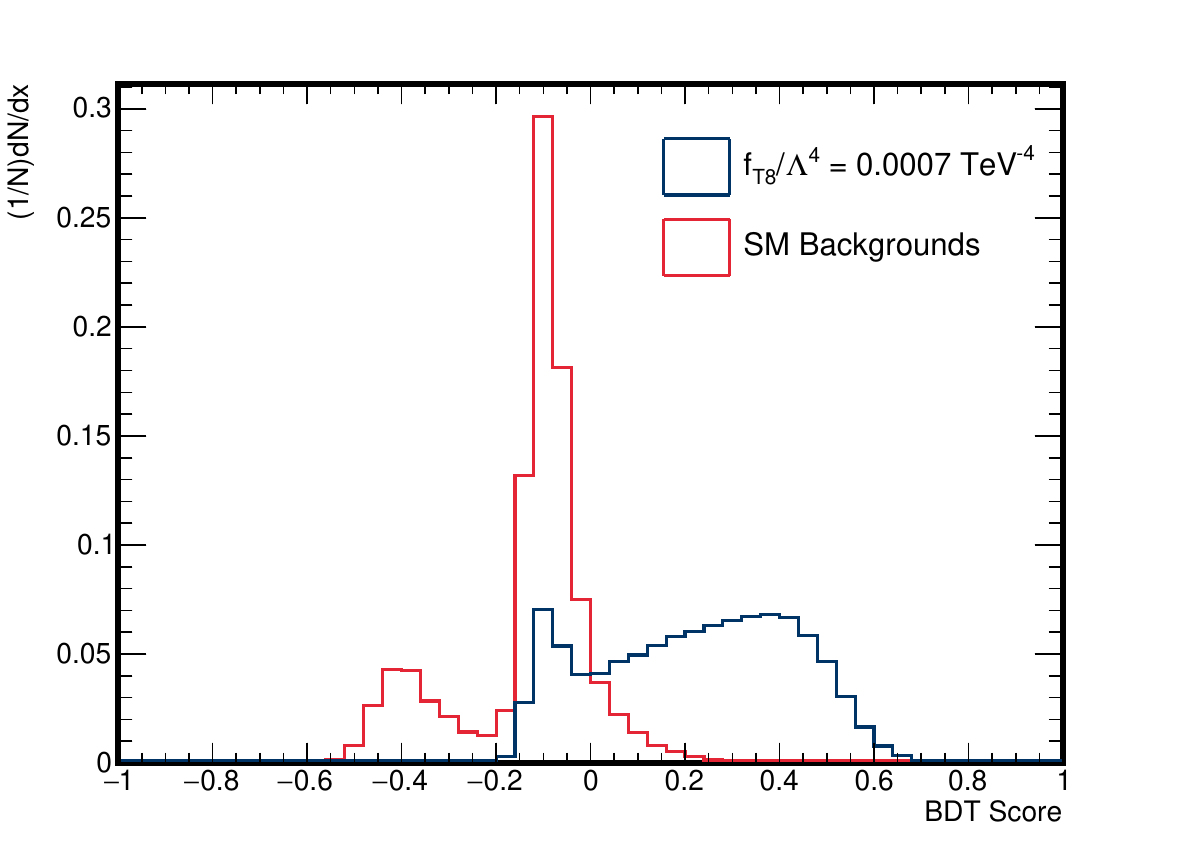}
\includegraphics[scale=0.40]{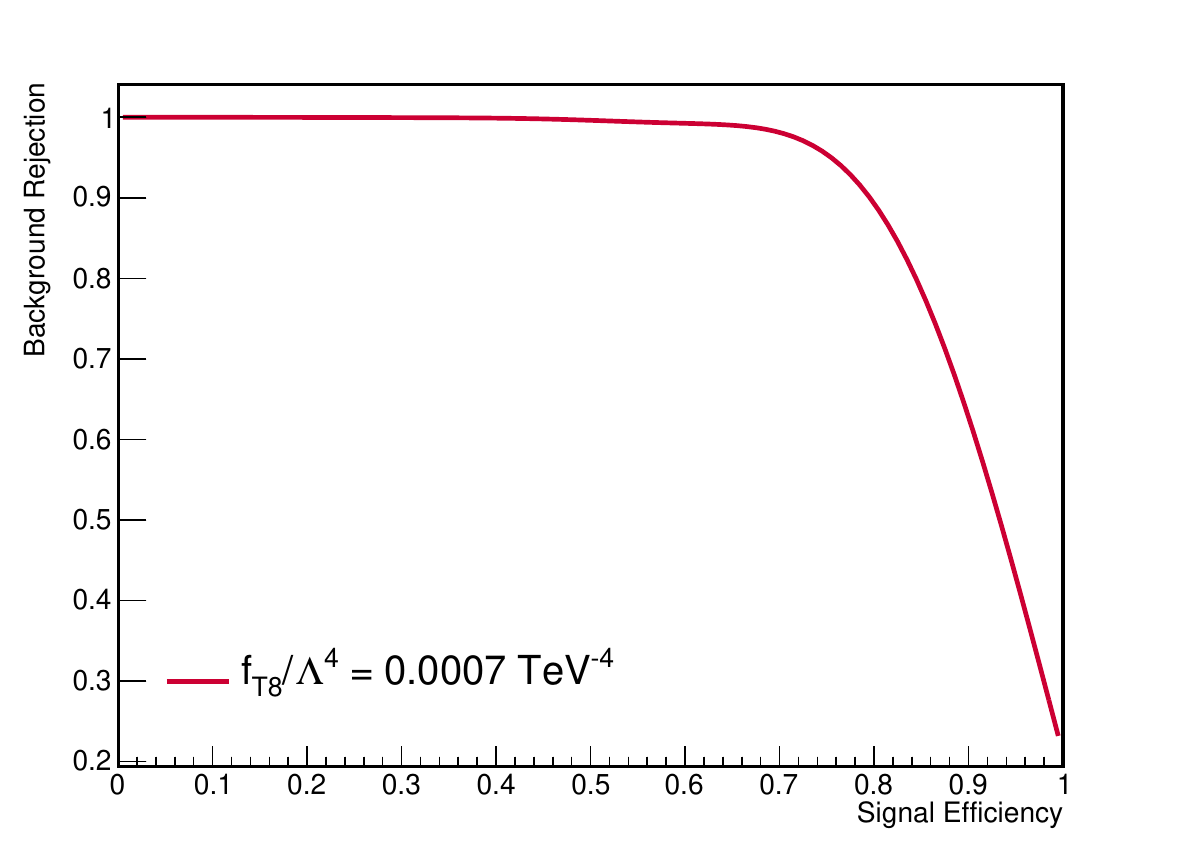}
\includegraphics[scale=0.40]{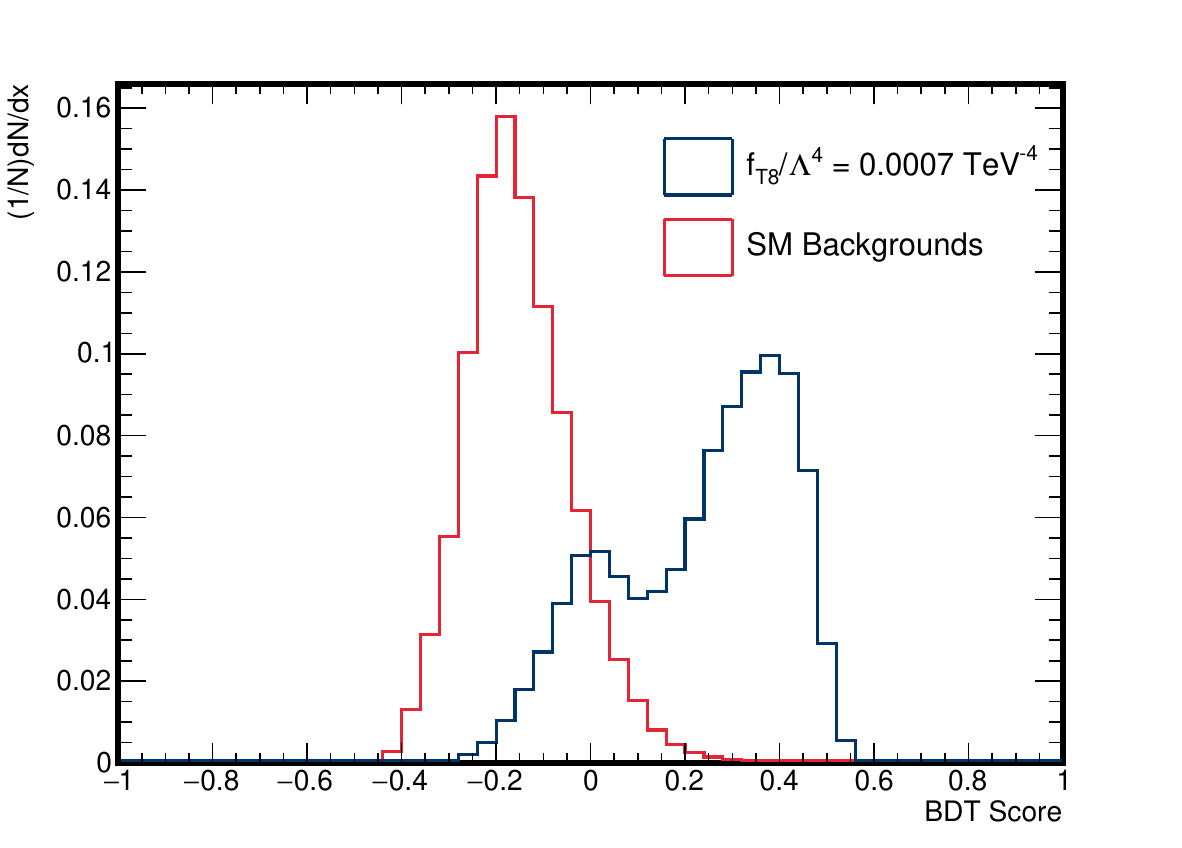}
\includegraphics[scale=0.40]{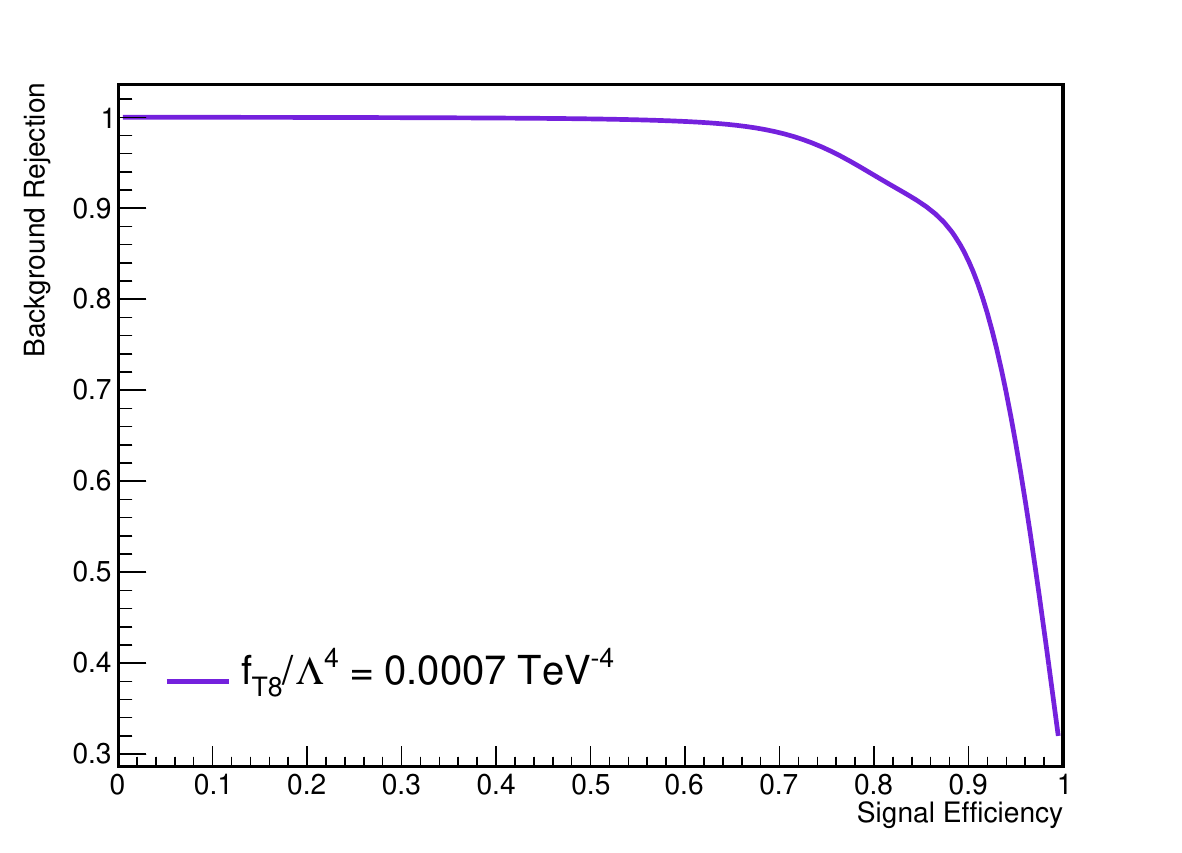}
\caption{Normalized BDT response distributions (left panels) and the corresponding Receiver Operating Characteristic (ROC) curves (right panels) for the EFT-induced signal
($f_{T8}/\Lambda^{4}=0.0007~\mathrm{TeV}^{-4}$) and the total Standard Model
background at $\sqrt{s}=10$ TeV. The upper (lower) panels correspond to the
$\proaa$
($\proza$) process.}
\label{fig:placeholder}
\end{figure}

The performance of the multivariate analysis based on the Boosted Decision Tree algorithm is evaluated for both signal processes at a center-of-mass energy of $\sqrt{s}=10$ TeV. The resulting BDT response distributions for the signal benchmark point $f_{T8}/\Lambda^{4}=0.0007~\mathrm{TeV}^{-4}$ and the corresponding Standard Model background contributions are presented in Fig.~\ref{fig:placeholder}.
As can be observed from the BDT score distributions, a clear separation between the signal and background hypotheses is achieved for both $\proaa$ and $\proza$ processes. The signal events tend to populate the higher BDT score region, whereas the Standard Model background is predominantly concentrated at lower score values. This behavior indicates that the kinematic and topological observables selected in the TMVA framework efficiently capture the characteristic features of EFT-induced anomalous interactions.
The corresponding Receiver Operating Characteristic (ROC) curves are also shown in Fig.~\ref{fig:placeholder}, where the background rejection efficiency is plotted as a function of the signal efficiency. In both processes, the ROC curves exhibit a strong background rejection performance across a broad range of signal efficiencies, demonstrating the strong discriminating power of the trained BDT classifier. In particular, high background rejection rates can be maintained even for relatively large signal efficiencies, which is essential for enhancing the sensitivity to dimension-8 anomalous quartic gauge couplings in the presence of sizable Standard Model contributions.
These results confirm that the employed multivariate strategy significantly improves the signal-to-background separation at $\sqrt{s}=10$ TeV and contributes to improve the expected sensitivity of a future muon collider to neutral anomalous quartic gauge interactions.

\begin{figure}[httb!]
\includegraphics[scale=0.40]{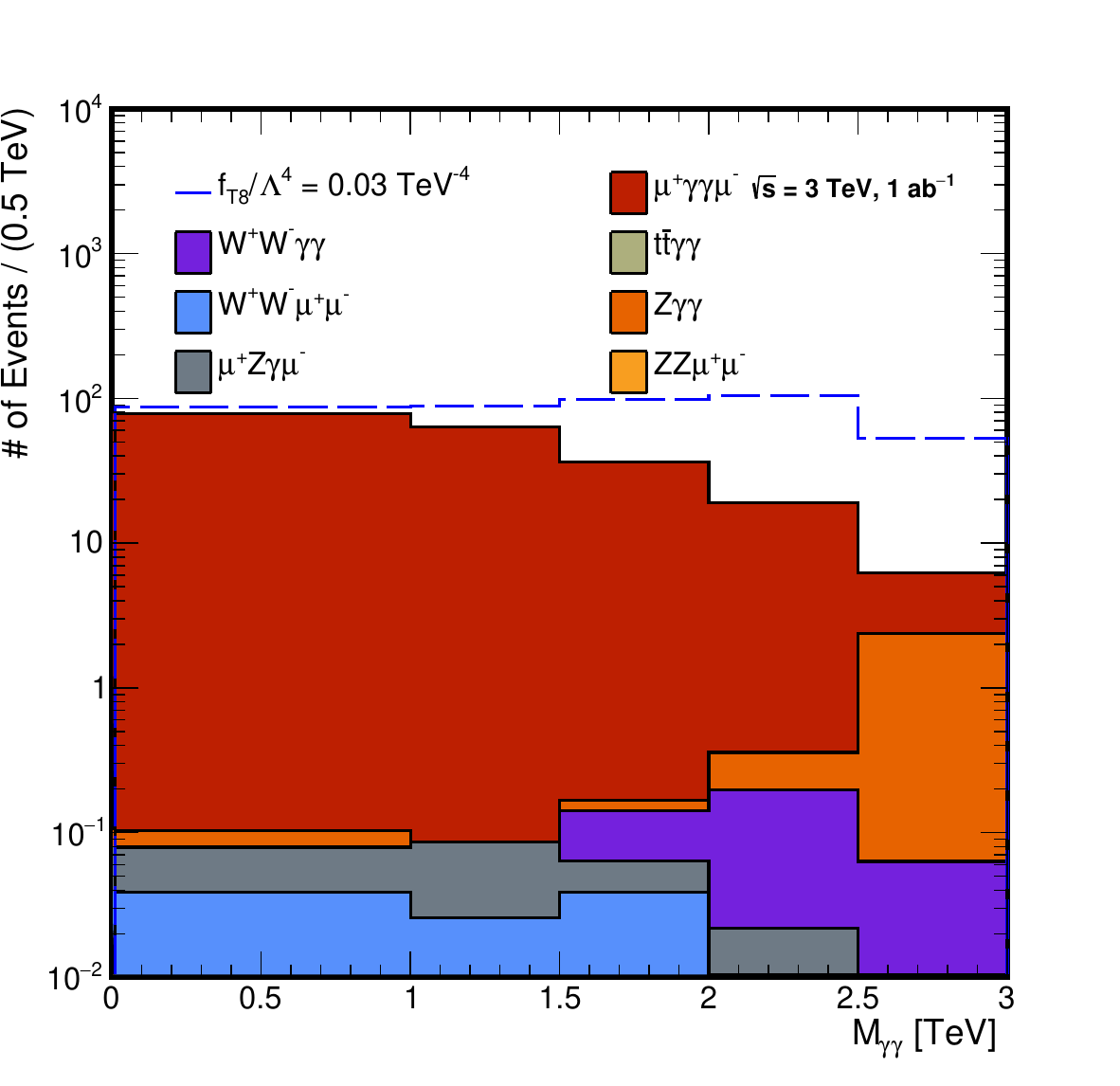}
\includegraphics[scale=0.40]{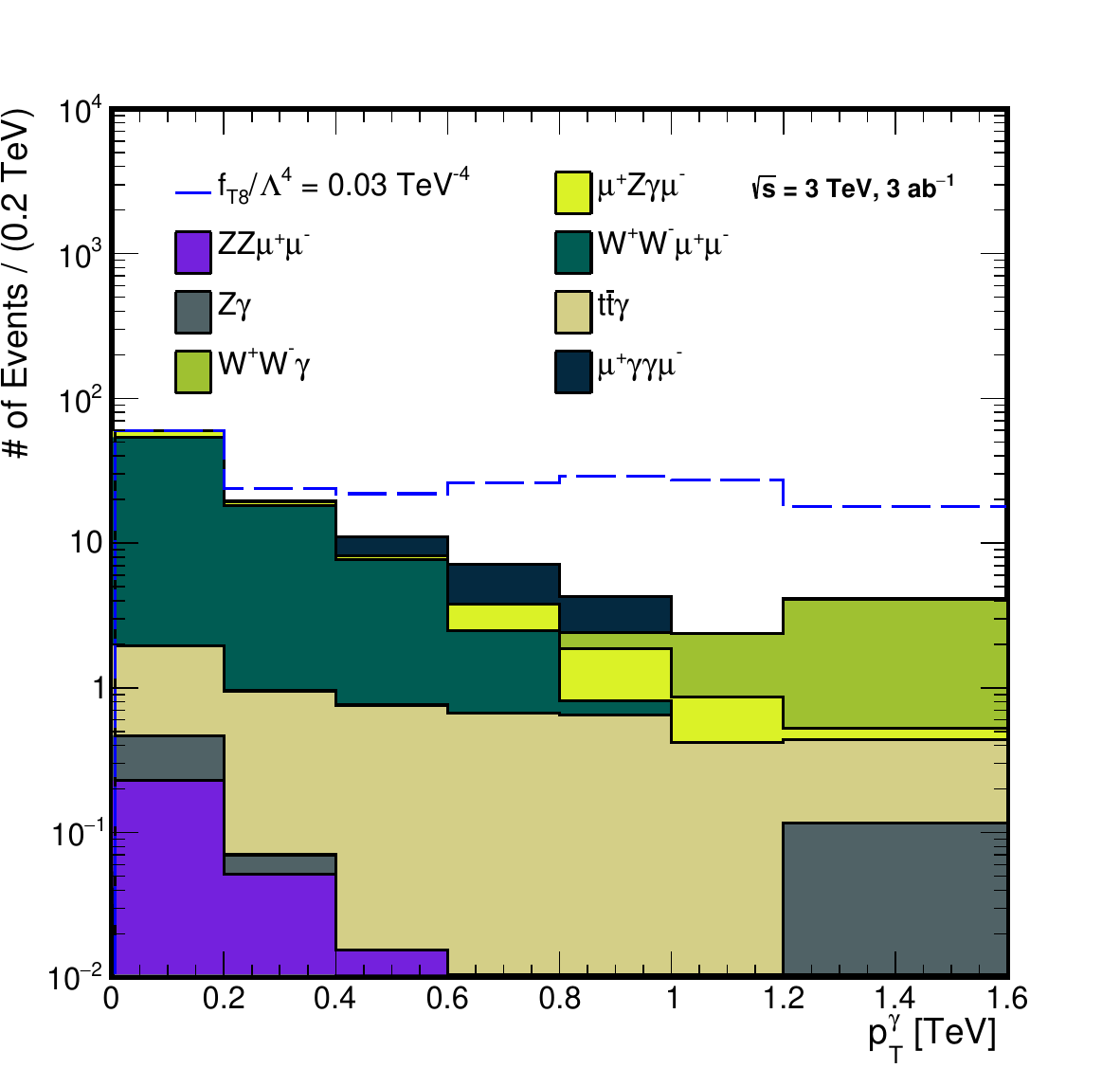}
\includegraphics[scale=0.40]{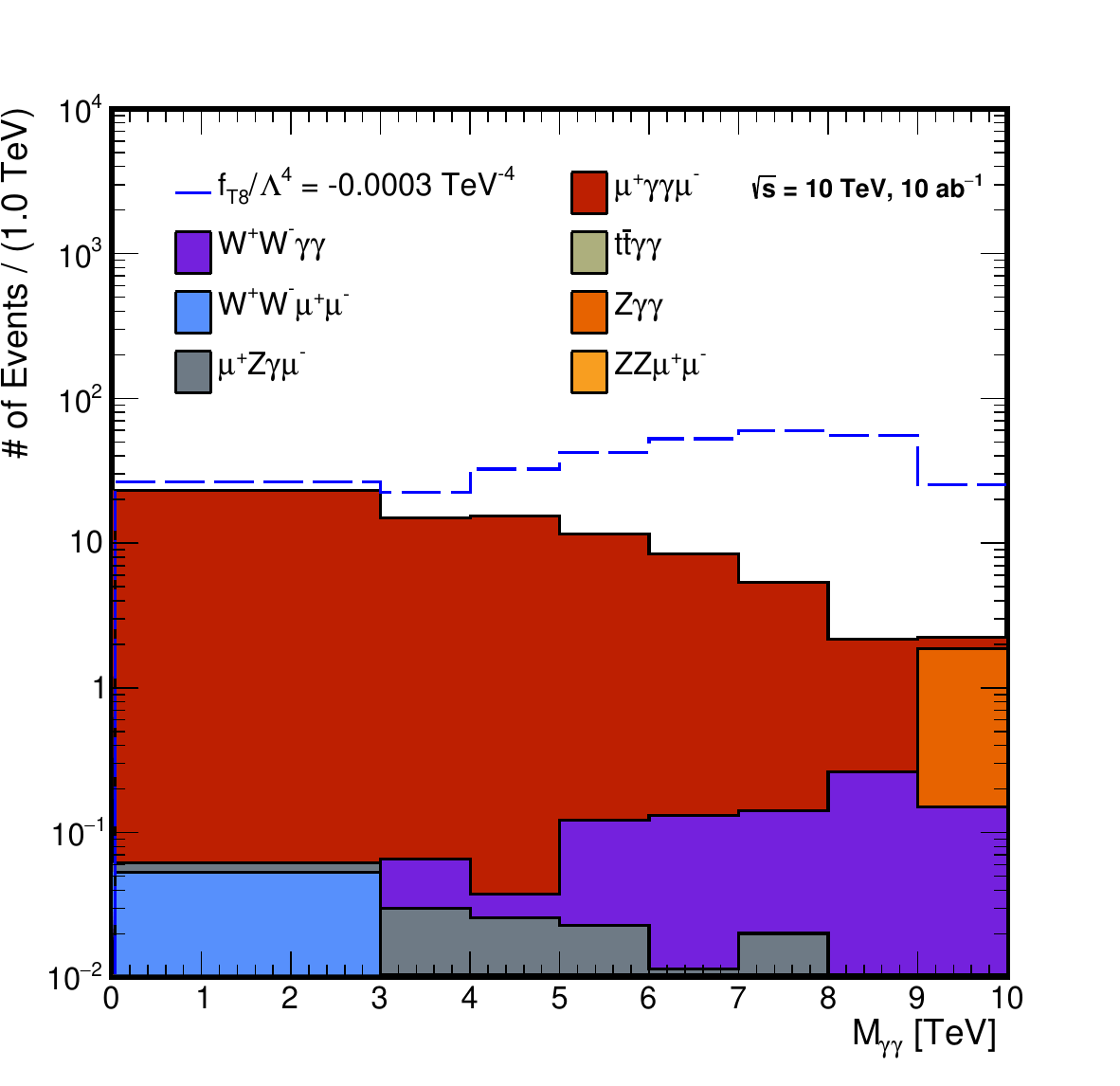}
\includegraphics[scale=0.40]{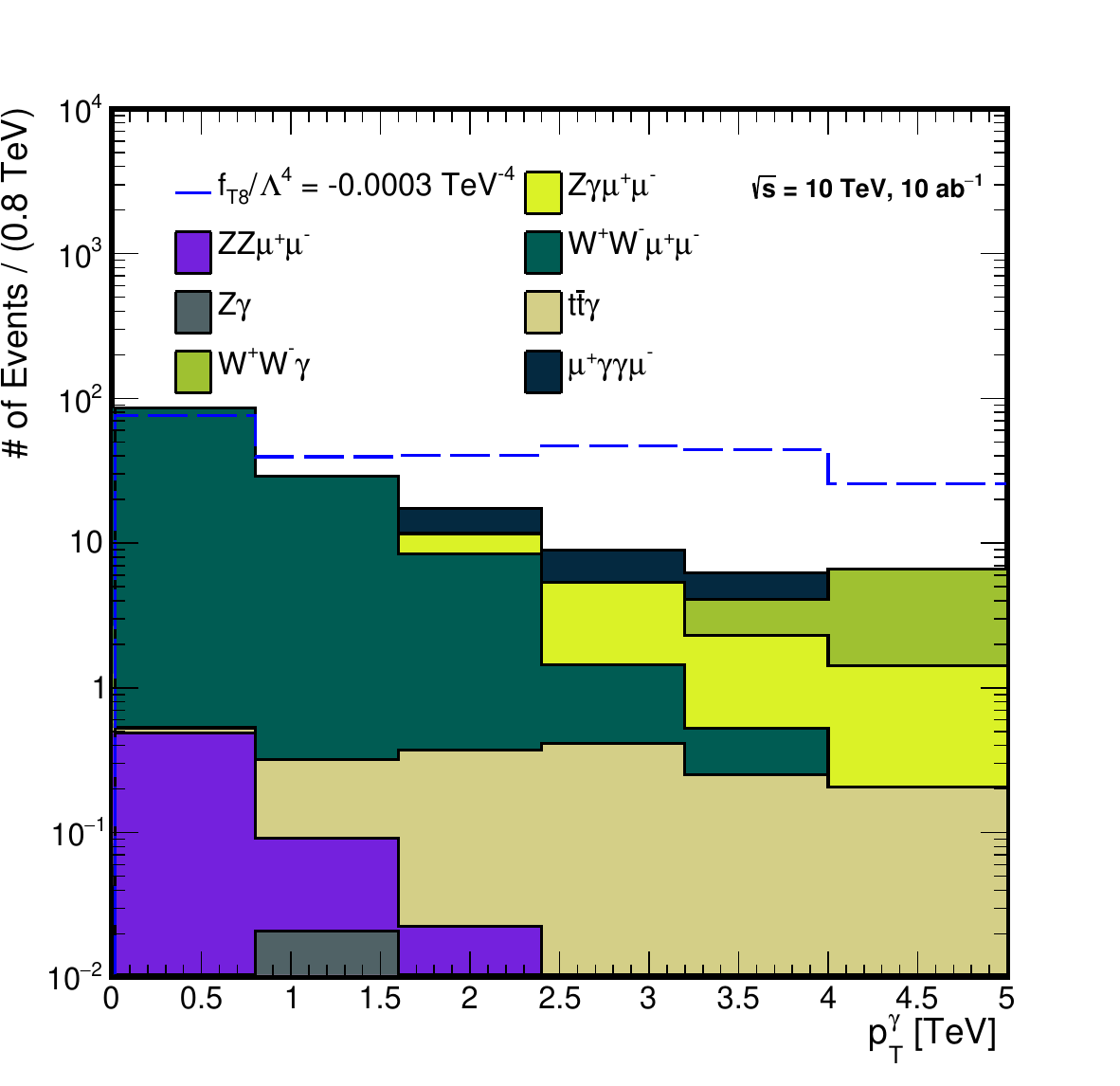}

\caption{The normalized invariant mass ($M_{\gamma \gamma}$) distribution of the two-photon system for the $\proaa$ process (left column) and the normalized transverse momentum ($p_{T}^{\gamma}$) distribution of the leading photon for the $\proza$ process (right column) at $\sqrt{s}=$3 TeV (Top row) and $\sqrt{s}=$10 TeV (Bottom row).\label{maa_invt_pt_ft8} }
\end{figure}

After imposing the corresponding BDT score requirements, the normalized invariant mass ($M_{\gamma\gamma}$) distribution of the di-photon system for the $\proaa$ process is shown in the left panel of Fig.~\ref{maa_invt_pt_ft8} , while the right panel displays the transverse momentum distribution of the leading photon for the 
$\proza$ process. The upper (lower) row corresponds to $\sqrt{s}=3$ TeV ($\sqrt{s}=10$ TeV) center-of-mass energy.

It is observed that, once the BDT-based selection is imposed on top of the basic and kinematic cuts, the SM background contributions are significantly reduced over a wide kinematic range, whereas the EFT signal exhibits a relatively enhanced population in the high-energy tails of the distributions. In particular, the $M_{\gamma\gamma}$ spectrum for the 
$\proaa$ process shows a pronounced excess in the multi-TeV region at $\sqrt{s}=10$ TeV, reflecting the characteristic energy-growing behaviour of dimension-8 operator contributions.

A similar trend is observed in the $p_T^\gamma$ distribution of the leading photon for the 
$\proza$ process, where the EFT-induced deviations become increasingly visible at higher transverse momentum values after the multivariate selection. This behaviour confirms that the combination of optimized kinematic requirements and BDT score cuts efficiently isolates the phase-space regions where anomalous quartic gauge couplings yield the largest impact, thereby improving the overall sensitivity of the analysis to the considered EFT operators.

\section{Sensitivities of Anomalous Quartic Gauge Couplings}\label{secIV}

\begin{figure}[httb!]
 \includegraphics[scale=0.40]{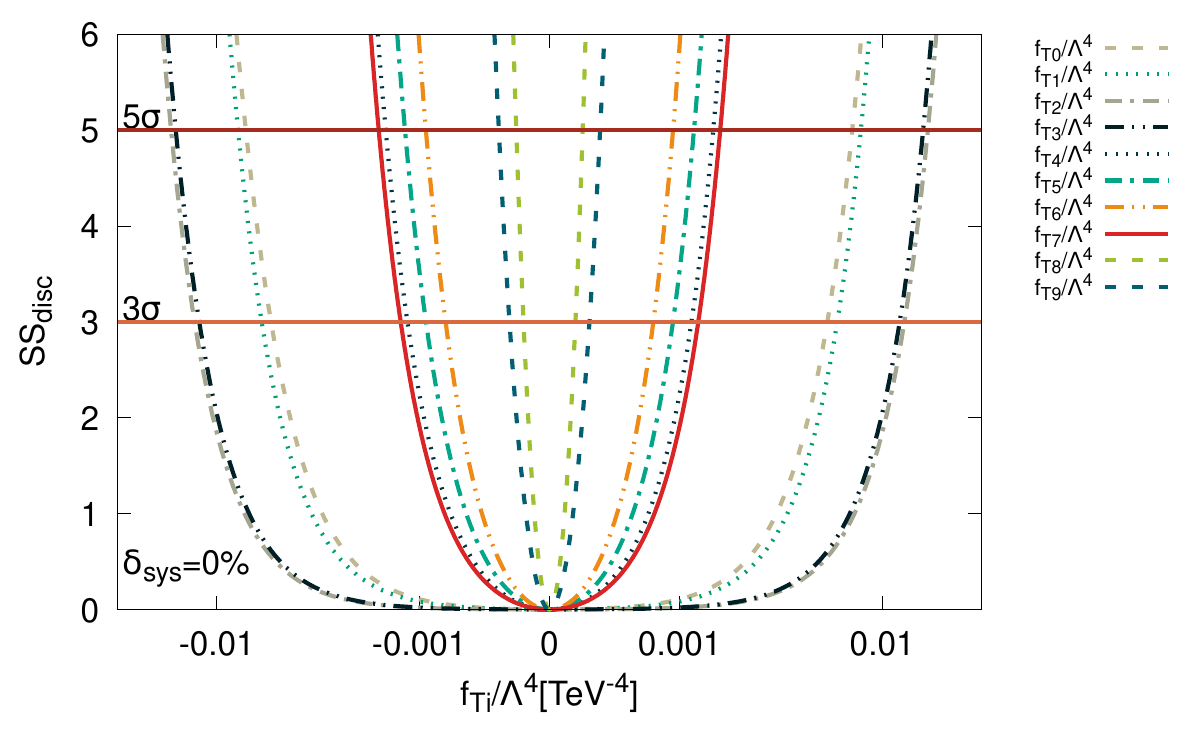}
 \includegraphics[scale=0.40]{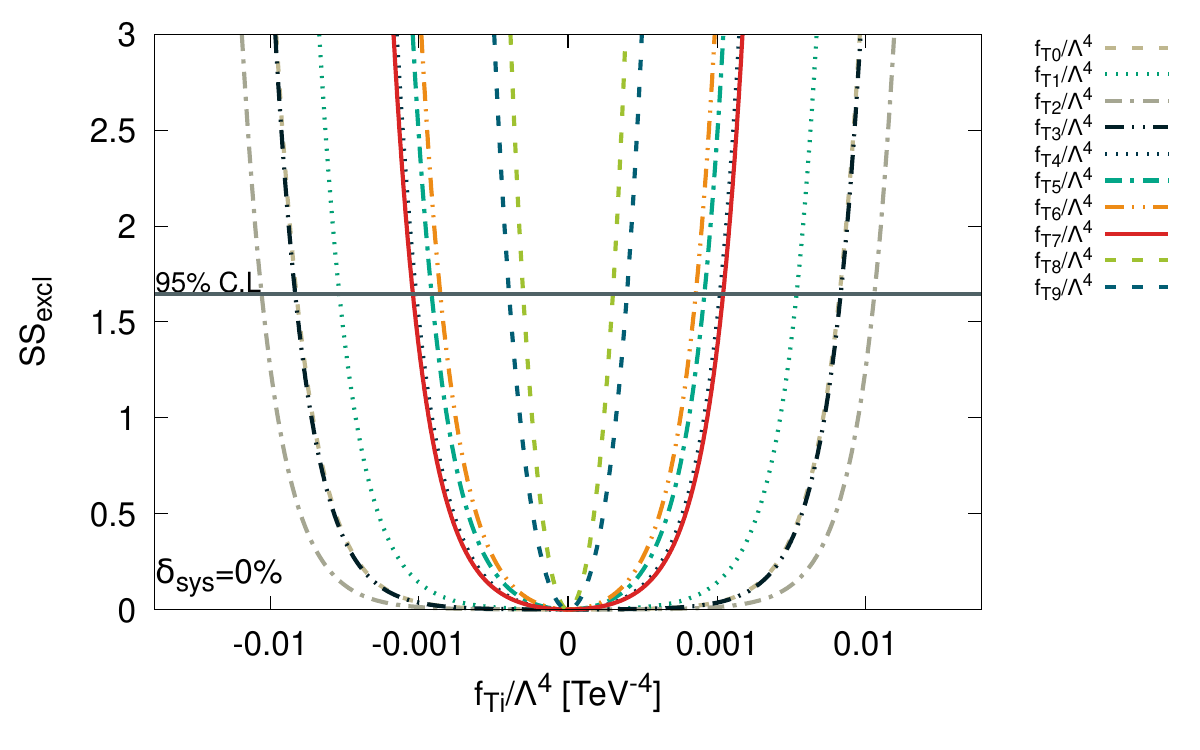}
 \includegraphics[scale=0.40]{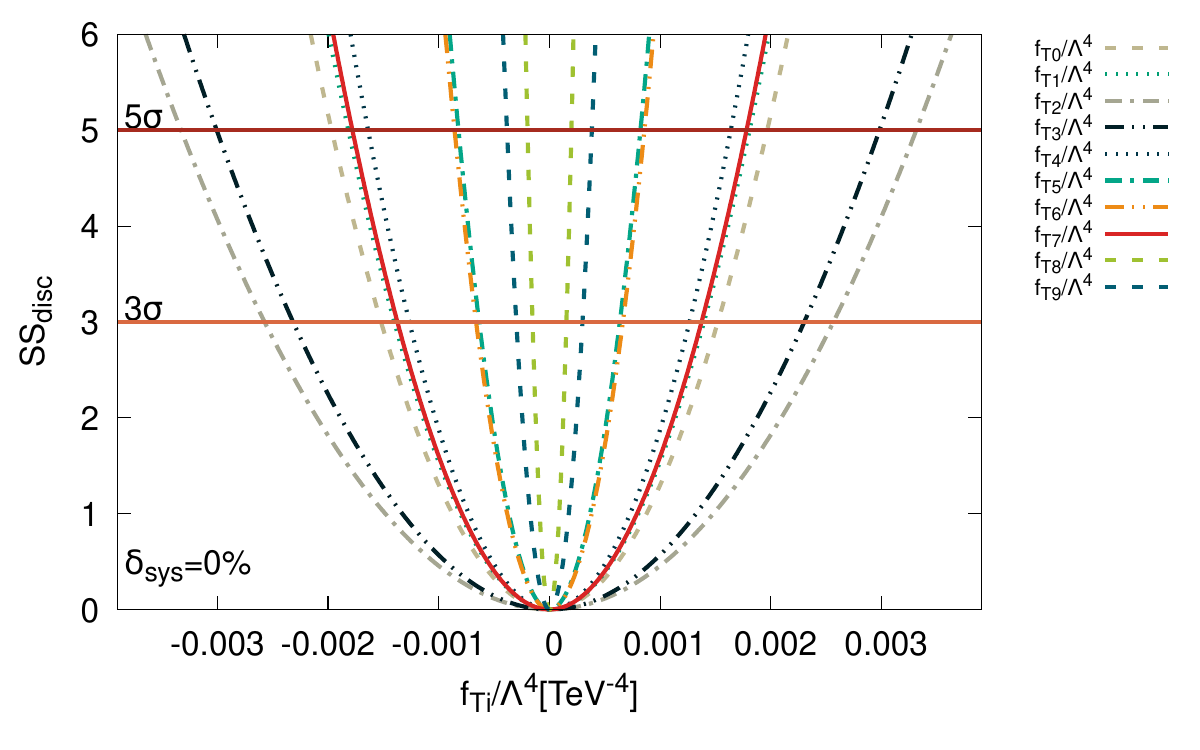}
 \includegraphics[scale=0.40]{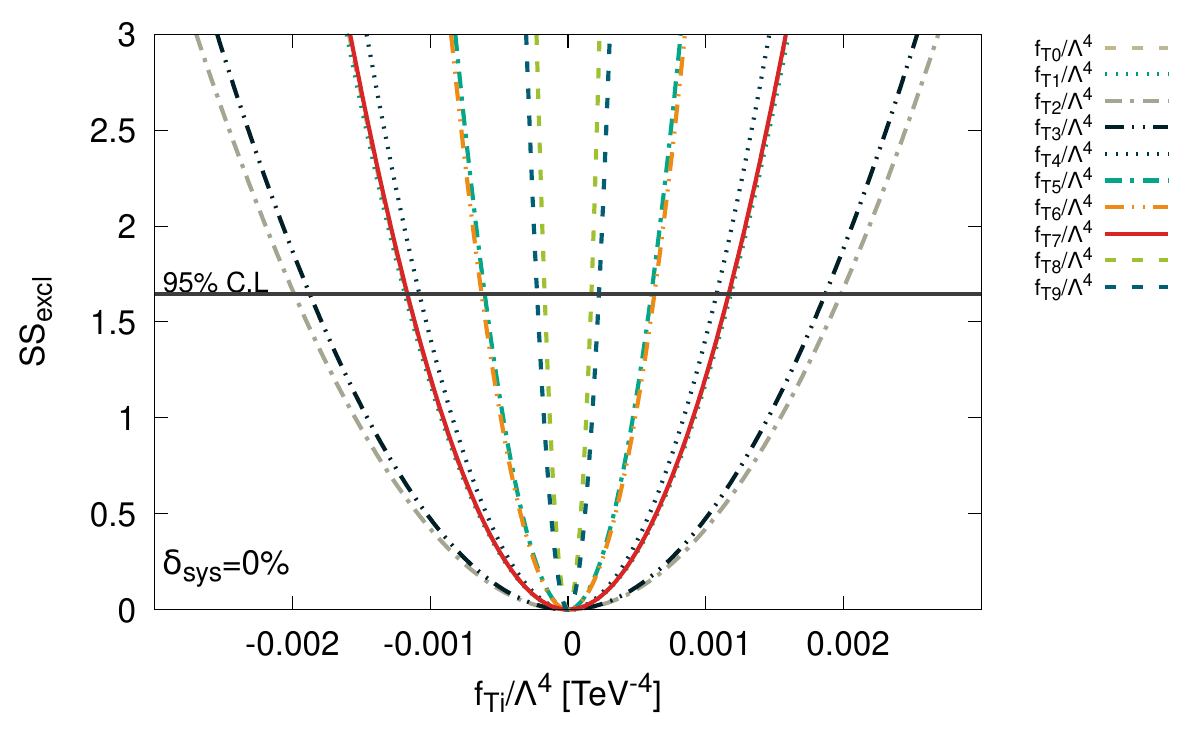}
\caption{The $SS_{disc}$ and $SS_{excl}$ as a function of anomalous couplings $f_{Ti}$ for the processes $\proaa$ (top row) and $\proza$ (bottom row) at $\com = 10$ TeV Muon Collider assuming no systematic uncertainties.}
\label{fig:ssdisc-excl}
\end{figure}

\begin{table}[httb!]
\caption{Obtained discovery ($3\sigma$ and $5\sigma$) and exclusion (95\%C.L) limits on the aQGC, $f_{Ti}/\Lambda^{4}$, without systematic uncertainty for the processes $\proaa$ and $\proza$ at 3 TeV and 10 TeV center of mass energy for Muon Collider.}
\label{tab:alllimits_nosys}
\begin{ruledtabular}
\begin{tabular}{l c |cc | cc}
 &  & \multicolumn{2}{c|}{$\sqrt{s}=3$ TeV MuCol}
    & \multicolumn{2}{c}{$\sqrt{s}=10$ TeV MuCol} \\
\hline
Coefficient & 
 & $\proaa$ & $\proza$
 & $\proaa$ & $\proza$ \\
\hline
\multirow{2}{*}{$f_{T0}/\Lambda^{4}$}
 & $3\sigma$ & [-7.00 ; 6.67]$\times10^{-1}$  & [-1.69 ; 1.69]$\times10^{-1}$  & [-1.15 ; 1.15]$\times10^{-2}$  & [-1.52 ; 1.52]$\times10^{-3}$  \\
 & $5\sigma$ & [-9.21 ; 9.20]$\times10^{-1}$ & [-2.19 ; 2.19]$\times10^{-1}$  & [-1.50 ; 1.50]$\times10^{-2}$ & [-1.97 ; 1.97]$\times10^{-3}$  \\
 & 95\%C.L & [-5.02 ; 5.02]$\times10^{-1}$  & [-1.39 ; 1.39]$\times10^{-1}$  & [-6.91 ; 6.91]$\times10^{-3}$ & [-1.17 ; 1.17]$\times10^{-3}$  \\
\hline
\multirow{2}{*}{$f_{T1}/\Lambda^{4}$}
 & $3\sigma$ & [-5.06 ; 4.83]$\times10^{-1}$ & [-1.64 ; 1.63]$\times10^{-1}$  & [-6.24 ; 6.24]$\times10^{-3}$  & [-1.40 ; 1.40]$\times10^{-3}$  \\
 & $5\sigma$ & [-6.80 ; 6.35]$\times10^{-1}$ & [-2.12 ; 2.12]$\times10^{-1}$ & [-8.10 ; 8.10]$\times10^{-3}$  & [-1.82 ; 1.81]$\times10^{-3}$  \\
 & 95\%C.L & [-3.91 ; 3.81]$\times10^{-1}$ &  [-1.35 ; 1.35]$\times10^{-1}$ & [-5.23 ; 5.23]$\times10^{-3}$ & [-1.19 ; 1.18]$\times10^{-3}$ \\
\hline
\multirow{2}{*}{$f_{T2}/\Lambda^{4}$}
 & $3\sigma$ & [-1.28 ; 1.28]$\times10^{0}$ & [-3.06 ; 3.04]$\times10^{-1}$  & [-1.28 ; 1.27]$\times10^{-2}$  &  [-2.57 ; 2.57]$\times10^{-3}$ \\
 & $5\sigma$ & [-1.67 ; 1.67]$\times10^{0}$ & [-3.96 ; 3.93]$\times10^{-1}$ & [-1.66 ; 1.65]$\times10^{-2}$ & [-3.32 ;3.31]$\times10^{-3}$ \\
 & 95\%C.L & [-1.02 ; 1.02]$\times10^{0}$ & [-2.36 ; 2.36]$\times10^{-1}$ & [-1.15 ; 1.15]$\times10^{-2}$ & [-1.99 ; 1.98]$\times10^{-3}$ \\
\hline
\multirow{2}{*}{$f_{T3}/\Lambda^{4}$}
 & $3\sigma$ & [-1.01 ; 1.01]$\times10^{0}$ & [-2.91 ; 2.90]$\times10^{-1}$ & [-8.63 ; 8.60]$\times10^{-3}$  & [-2.30 ; 2.31]$\times10^{-3}$  \\
 & $5\sigma$ & [-1.33 ; 1.31]$\times10^{0}$ & [-3.77 ; 3.75]$\times10^{-1}$  & [-1.13 ; 1.12]$\times10^{-2}$  & [-3.00 ; 2.98]$\times10^{-3}$ \\
 & 95\%C.L & [-8.02 ; 7.98]$\times10^{-1}$ & [-2.28 ; 2.27]$\times10^{-1}$ & [-6.83 ; 6.81]$\times10^{-3}$ & [-1.87 ; 1.87]$\times10^{-3}$ \\
\hline
\multirow{2}{*}{$f_{T4}/\Lambda^{4}$}
 & $3\sigma$ & [-1.40 ; 1.40]$\times10^{-1}$ & [-1.57 ; 1.57]$\times10^{-1}$  & [-1.16 ; 1.16]$\times10^{-3}$  & [-1.26 ; 1.26]$\times10^{-3}$  \\
 & $5\sigma$ & [-1.82 ; 1.82]$\times10^{-1}$ & [-2.03 ; 2.03]$\times10^{-1}$  & [-1.49 ; 1.49]$\times10^{-3}$  & [-1.64 ; 1.64]$\times10^{-3}$  \\
 & 95\%C.L & [-1.26 ; 1.26]$\times10^{-1}$ & [-1.34 ; 1.34]$\times10^{-1}$ & [-1.04 ; 1.04]$\times10^{-3}$ & [-1.09 ; 1.08]$\times10^{-3}$ \\
\hline
\multirow{2}{*}{$f_{T5}/\Lambda^{4}$}
 & $3\sigma$ & [-1.04 ; 1.04]$\times10^{-1}$ & [-7.66 ; 7.66]$\times10^{-2}$ & [-8.10 ; 7.94]$\times10^{-4}$  & [-6.35 ; 6.35]$\times10^{-4}$  \\
 & $5\sigma$ & [-1.35 ; 1.35]$\times10^{-1}$ & [-9.91 ; 9.91]$\times10^{-2}$ & [-1.05 ; 1.02]$\times10^{-3}$  & [-8.22 ; 8.22]$\times10^{-4}$  \\
 & 95\%C.L & [-9.41 ; 9.41]$\times10^{-2}$ & [-6.79; 6.79]$\times10^{-2}$ & [-6.59 ; 6.52]$\times10^{-4}$ & [-6.05 ; 6.05]$\times10^{-4}$ \\
\hline
\multirow{2}{*}{$f_{T6}/\Lambda^{4}$}
 & $3\sigma$ & [-7.15 ; 7.15]$\times10^{-2}$ & [-8.21 ; 8.21]$\times10^{-2}$  & [-7.12 ; 7.12]$\times10^{-4}$  & [-7.31 ; 7.31]$\times10^{-4}$ \\
 & $5\sigma$ & [-9.25 ; 9.25]$\times10^{-2}$ & [-1.06 ; 1.06]$\times10^{-1}$  & [-9.20 ; 9.20]$\times10^{-4}$  & [-9.44 ; 9.44]$\times10^{-4}$ \\
 & 95\%C.L & [-6.52 ; 6.52]$\times10^{-2}$& [-7.18 ; 7.18]$\times10^{-2}$ &[-7.09 ; 7.09]$\times10^{-4}$ &[-5.06 ; 5.06]$\times10^{-4}$ \\
\hline
\multirow{2}{*}{$f_{T7}/\Lambda^{4}$}
 & $3\sigma$ & [-1.49 ; 1.49]$\times10^{-1}$ & [-1.71 ; 1.71]$\times10^{-1}$ & [-1.25 ; 1.25]$\times10^{-3}$ & [-1.37 ; 1.37]$\times10^{-3}$ \\
 & $5\sigma$ & [-1.93 ; 1.93]$\times10^{-1}$ & [-2.21 ; 2.21]$\times10^{-1}$ & [-1.62 ; 1.62]$\times10^{-3}$  & [-1.78 ; 1.78]$\times10^{-3}$  \\
 & 95\%C.L & [-1.32 ; 1.32]$\times10^{-1}$ & [-1.44 ; 1.44]$\times10^{-1}$ & [-1.09 ; 1.09]$\times10^{-3}$ & [-1.16 ; 1.16]$\times10^{-3}$ \\
\hline
\multirow{2}{*}{$f_{T8}/\Lambda^{4}$}
 & $3\sigma$ & [-1.30 ; 1.30]$\times10^{-2}$ & [-1.71 ; 1.71]$\times10^{-2}$  & [-1.15 ; 1.15]$\times10^{-4}$  & [-1.53 ; 1.53]$\times10^{-4}$  \\
 & $5\sigma$ & [-1.69 ; 1.69]$\times10^{-2}$ & [-2.20 ; 2.20]$\times10^{-2} $ & [-1.49 ; 1.49]$\times10^{-4}$ & [-1.98 ; 1.98]$\times10^{-4}$  \\
 & 95\%C.L & [-1.17 ; 1.17]$\times10^{-2}$ & [-1.47 ; 1.47]$\times10^{-2}$ & [-1.06 ; 1.06]$\times10^{-4}$ & [-1.13 ; 1.13]$\times10^{-4}$ \\
\hline
\multirow{2}{*}{$f_{T9}/\Lambda^{4}$}
 & $3\sigma$ & [-2.52 ; 2.52]$\times10^{-2}$ & [-3.81 ; 3.81]$\times10^{-2}$ & [-2.14 ; 2.14]$\times10^{-4}$ & [-3.00 ; 3.00]$\times10^{-4}$ \\
 & $5\sigma$ & [-3.27 ; 3.27]$\times10^{-2}$ & [-4.85 ; 4.85]$\times10^{-2}$ & [-2.78 ; 2.78]$\times10^{-4}$  & [-3.84 ; 3.84]$\times10^{-4}$ \\
 & 95\%C.L & [-2.13 ; 2.13]$\times10^{-2}$ & [-2.82 ; 2.82]$\times10^{-2}$ & [-1.73 ; 1.73]$\times10^{-4}$ & [-2.25 ; 2.24]$\times10^{-4}$ \\
\hline
\hline
\end{tabular}
\end{ruledtabular}
\end{table}

\begin{figure}[httb!]
\includegraphics[scale=0.8]{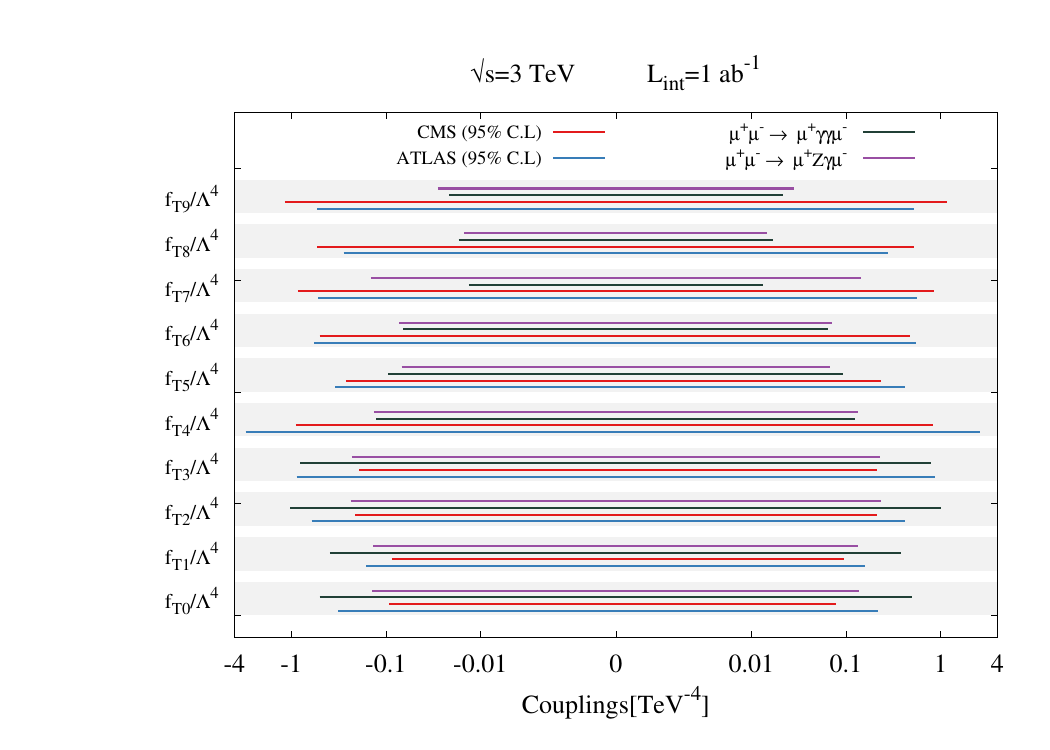}
\includegraphics[scale=0.8]{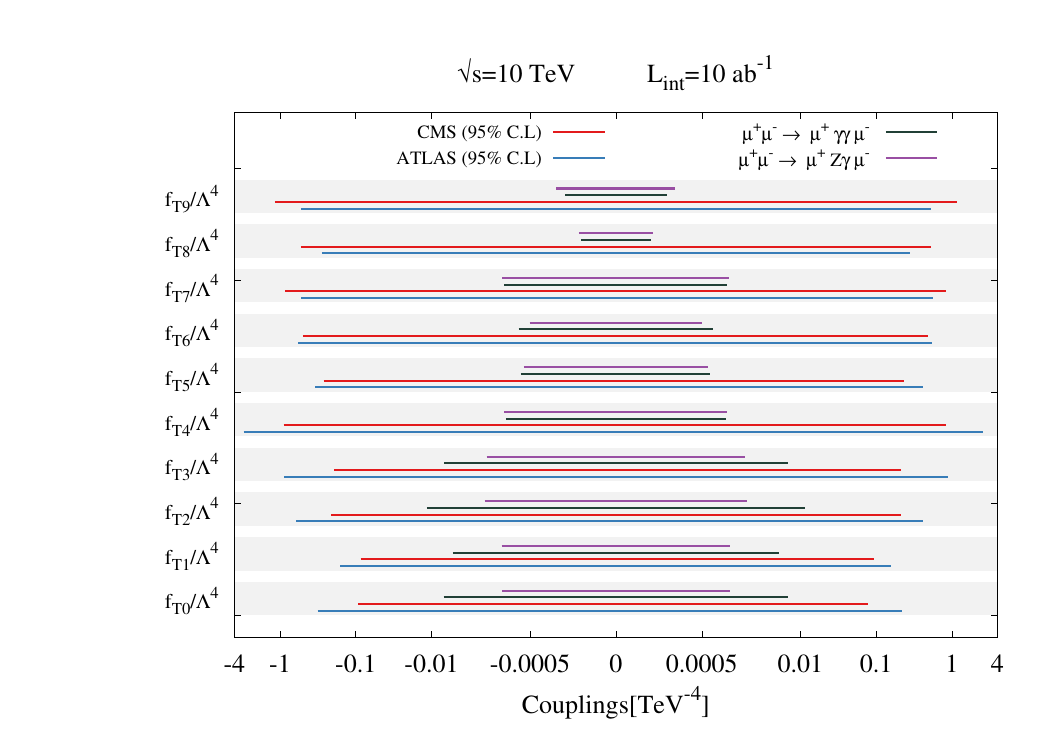}
\caption{Comparison of the most recent experimental limits reported by the ATLAS and CMS Collaborations and the obtained 95\% C.L limits and our obtained 95\%C.L limits on the aQGC, $f_{Ti}/\Lambda^{4}$, without systematic uncertainty for the processes $\proaa$ and $\proza$ at 3 TeV (top) and 10 TeV (bottom) center of mass energy for Muon Collider.}
\label{fig:limits}
\end{figure}

In this section, we present the projected discovery and exclusion sensitivities 
for the anomalous quartic gauge couplings $f_{Ti}/\Lambda^{4}$ in the 
$\proaa$ and $\proza$ processes at the $\sqrt{s}=3$ TeV ($\mathcal{L}_{\mathrm{int}}=1~\mathrm{ab}^{-1}$) and 
$\sqrt{s}=10$ TeV ($\mathcal{L}_{\mathrm{int}}=10~\mathrm{ab}^{-1}$) Muon Collider stages.

Following the application of the baseline event selection and the multivariate 
classification based on the BDT response, the signal ($S$) and total Standard 
Model background ($B$) yields are extracted from the optimized high–energy 
regions of the kinematic observables that exhibit the strongest EFT sensitivity. 
In particular, the event counting is performed using the invariant mass of the 
two–photon system ($M_{\gamma\gamma}$) for the $\proaa$ process and the transverse 
momentum of the leading photon ($p_T^{\gamma}$) for the $\proza$ process, after the BDT score selection.

The median expected statistical significances for discovery 
($\mathcal{SS}_{\mathrm{disc}}$) and exclusion 
($\mathcal{SS}_{\mathrm{excl}}$), including the effect of systematic 
uncertainties $\delta$, are computed using the profile likelihood–based 
expressions~\cite{Cowan:2010js,Kumar:2015tna}:

\begin{equation}
\mathcal{SS}_{\text{disc}} = \sqrt{2 \left[ (S+B) \ln \left (\frac{(S+B)(1 + \delta^{2}B)}{B + \delta^{2} B (S+B)} \right)  -\frac{1}{\delta^{2}} \ln \left(1 + \delta^{2} \frac{S}{1 + \delta^{2}B} \right) \right] }
\end{equation}
\begin{equation}
\mathcal{SS}_{\text{excl}} = \sqrt{2 \left[S - B \ln \left( \frac{B+S+x}{2B}\right) -\frac{1}{\delta^{2}} \ln \left ( \frac{B-S+x}{2B}\right) \right] - (B+S-x)(1 + \frac{1}{\delta^{2}B})}
\end{equation}
where $x = \sqrt{(S + B)^{2} - 4\delta^{2}SB^{2}/(1 + \delta^{2}B)}$, with $S$ and $B$ denoting the expected number of signal and total background events, 
respectively. In the limit of vanishing systematic uncertainty 
($\delta\rightarrow 0$), the above expressions reduce to
\begin{equation}
\mathcal{SS}_{\text{disc}} = \sqrt{2 \left[(S + B) \ln \left ( 1 + S/B \right) - S \right] }
\end{equation}
\begin{equation}
\mathcal{SS}_{\text{excl}} = \sqrt{2 \left[S - B \ln \left ( 1 + S/B \right) \right]}
\end{equation}
Regions satisfying $\mathcal{SS}_{\mathrm{disc}}\geq 3\sigma$ 
($5\sigma$) are interpreted as evidence (discovery) reach, whereas 
$\mathcal{SS}_{\mathrm{excl}}\leq 1.645$ corresponds to exclusion at the 
$95\%$ confidence level.

The resulting statistical significances as functions of the anomalous EFT 
couplings are shown in Fig.~\ref{fig:ssdisc-excl} for the 
$\proaa$ (top panels) and 
$\proza$ (bottom panels) processes 
at $\sqrt{s}=10$ TeV in the absence of systematic uncertainties. The left 
(right) column illustrates the discovery (exclusion) sensitivity, where the 
horizontal reference lines indicate the $3\sigma$, $5\sigma$, and $95\%$ C.L. 
thresholds, respectively. The projected bounds on the anomalous quartic gauge 
couplings $f_{Ti}/\Lambda^{4}$ are obtained from the intersection points of the 
significance curves with these reference levels.

The projected sensitivity limits obtained in this analysis are summarized in 
Table~\ref{tab:alllimits_nosys}, where the discovery ($3\sigma$ and $5\sigma$) 
and exclusion ($95\%$ C.L.) bounds on the anomalous quartic gauge couplings 
\couplings are presented for the 
$\proaa$ and 
$\proza$ processes at 
$\sqrt{s}=3$ TeV with $\mathcal{L}_{\mathrm{int}}=1~\mathrm{ab}^{-1}$ and 
$\sqrt{s}=10$ TeV with $\mathcal{L}_{\mathrm{int}}=10~\mathrm{ab}^{-1}$.

A systematic improvement in the sensitivity reach is observed when 
increasing the center-of-mass energy from 3 TeV to 10 TeV. 
For most of the dimension–8 operators considered in this study, 
the limits derived from the 
$\proza$ process are found 
to be more stringent than those obtained from the 
$\proaa$ process. 
This behaviour can be attributed to the presence of the 
$Z\rightarrow\nu\bar{\nu}$ decay mode in the former, which introduces 
a genuine missing transverse energy signature that significantly 
suppresses fully visible Standard Model backgrounds. As a result, 
the signal–to–background separation achieved by the BDT classifier 
is improved, leading to enhanced ROC performance and consequently 
tighter projected EFT limits for the anomalous quartic gauge couplings.

The same process $\mu^+\mu^- \to \mu^+ Z\gamma \mu^-$ was analyzed using the $f_{T,i}/\Lambda^4$ parametrization at same the center-of-mass energy and integrated luminosity in Ref.~\cite{Gutierrez-Rodriguez:2025wcy}. In that study, the reported limits on the anomalous couplings are typically of the order of $\mathcal{O}(10^{-2})~\text{TeV}^{-4}$ in the absence of systematic uncertainties. In contrast, our results for the $\mu^+\mu^- \to \mu^+ Z\gamma \mu^-$ process at $\sqrt{s}=10~\text{TeV}$ (see Table VI) reach sensitivities at the level of $\mathcal{O}(10^{-3})$ and, for some operators, down to $\mathcal{O}(10^{-4})~\text{TeV}^{-4}$. This corresponds to an improvement of about one order of magnitude in general, and up to two orders of magnitude for specific operators such as $f_{T5}/\Lambda^4$, $f_{T6}/\Lambda^4$, $f_{T8}/\Lambda^4$ and $f_{T9}/\Lambda^4$. This improvement can be mainly attributed to the use of multivariate analysis techniques, in particular Boosted Decision Trees (BDT), together with optimized kinematical selections, which significantly enhance the signal-to-background discrimination. A similar level of improvement is also observed for the $\mu^+\mu^- \to \mu^+ \gamma\gamma \mu^-$ channel.

A comparison with the currently available one–dimensional limits from 
the ATLAS and CMS collaborations at $\sqrt{s}=13$ TeV, summarized in 
Table~II, demonstrates the substantial improvement achievable at a 
future muon collider. Existing LHC constraints on the 
\couplings operators are typically at the 
$\mathcal{O}(10^{-1})$--$\mathcal{O}(1)$ TeV$^{-4}$ level, depending on 
the production process and analysis strategy. The projected sensitivities 
at the 3 TeV Muon Collider already improve these bounds by approximately 
one order of magnitude for several operators.

The improvement becomes particularly pronounced in the 10 TeV scenario, 
where the sensitivity reach extends to the 
$\mathcal{O}(10^{-3})$--$\mathcal{O}(10^{-4})$ TeV$^{-4}$ level. This 
corresponds to an enhancement of up to two to three orders of magnitude 
with respect to the present LHC limits. The dominant gain originates 
from the strong energy dependence of dimension–8 EFT contributions, 
which exhibit an enhanced energy dependence within the EFT 
validity regime, together with the 
background–suppressed experimental environment provided by a 
high–energy lepton collider.

For a clearer comparison with existing experimental constraints, the 95\% C.L. limits listed in Table~\ref{tab:alllimits_nosys}  are illustrated in Fig.~\ref{fig:limits}. The upper panel corresponds to $\sqrt{s}=3$ TeV with $L_{\mathrm{int}}=1\,\mathrm{ab^{-1}}$, while the lower panel shows the $\sqrt{s}=10$ TeV and $L_{\mathrm{int}}=10\,\mathrm{ab^{-1}}$ scenario. In both panels, the red and blue lines represent the current CMS and ATLAS bounds, respectively, while the black lines denote the projected limits from the $\mu^+\gamma\gamma\mu^-$ process, whereas the purple lines correspond to the $\mu^+Z\gamma\mu^-$ process obtained in this analysis. The graphical representation makes the improvement in sensitivity immediately visible across all operators \couplings. The projected muon collider sensitivities clearly exceed the present LHC constraints, particularly in the $\sqrt{s}=10$ TeV scenario where the accessible parameter space is significantly reduced.


Systematic uncertainties play an important role in determining realistic
sensitivity limits at future collider experiments. These uncertainties may
originate from theoretical cross-section predictions, higher-order QCD and electroweak corrections, luminosity measurements, and particle misidentification effects. In this analysis, simulations are performed at leading-order accuracy, and higher-order corrections are not explicitly included. Their impact is expected to modify the overall normalization but not the qualitative sensitivity behavior presented in this work. Therefore, $\delta_{\mathrm{sys}}=10\%$ systematic uncertainty is considered to observe the impact of systematic uncertainties on the projected limits. The inclusion of a 10\% systematic uncertainty leads to a moderate relaxation of the bounds for all considered operators and processes, as expected. Nevertheless, the overall sensitivity remains largely preserved. This behavior indicates that the analysis is predominantly statistics-driven, and that systematic effects at the level considered here do not significantly degrade the discovery and exclusion potential. In particular, the $\sqrt{s}=10$ TeV muon collider scenario continues to provide strong constraints on the \couplings operators even in the presence of systematics.

Overall, these results indicate that the sensitivity is not significantly degraded 
for the assumed level of systematic uncertainty, further supporting the potential of future muon collider experiments for precision EFT studies.

\section{Conclusion}\label{conc}

In this work, we investigated the sensitivity of a future muon collider to
dimension-eight anomalous quartic gauge couplings parameterized in terms of
\couplings. The analysis was performed for the vector boson
scattering processes $\proaa$ and $\proza$ at center-of-mass energies of
3 TeV and 10 TeV, assuming integrated luminosities of
$1\,\mathrm{ab^{-1}}$ and $10\,\mathrm{ab^{-1}}$, respectively. Detector-level
effects were taken into account in order to obtain realistic sensitivity
projections for future collider experiments.

A multivariate analysis based on Boosted Decision Trees (BDT) was employed
to enhance the separation between signal and Standard Model backgrounds.
Kinematic observables associated with the final-state muons and photons,
together with reconstructed quantities such as the dilepton invariant mass
($M_{\mu_1\mu_2}$) and photon centrality variables, were found to provide
strong discrimination power. In particular, the di-photon centrality
($\gamma\gamma$-centrality) observable for the $\proaa$ process and the
photon centrality ($\gamma$-centrality) observable for the
$\proza$ process play a key role in the signal–background separation achieved by the BDT classifier. Furthermore, high-energy observables such as the normalized invariant
mass ($M_{\gamma\gamma}$) in the $\proaa$ process and the
normalized leading-photon transverse momentum ($p_{T}^{\gamma}$)
distribution in the $\proza$ process significantly improve the sensitivity to anomalous interactions.

The obtained sensitivity projections indicate that a high-energy muon
collider would substantially extend the current experimental reach on
anomalous quartic gauge couplings. A clear improvement in the attainable
limits is observed when increasing the center-of-mass energy from 3 TeV
to 10 TeV. For most of the operators considered in this study, the limits
derived from the $\proza$ process are more stringent than those obtained from the
$\proaa$ process. This enhancement originates from the presence of the
$Z\rightarrow\nu\bar{\nu}$ decay mode in the former, which introduces
a genuine missing transverse energy signature that efficiently suppresses
fully visible Standard Model backgrounds. Consequently, the signal–background separation achieved in the BDT analysis is improved, leading to tighter projected limits on the \couplings operators.

Overall, these results highlight the strong potential of future muon
collider programs for precision studies of the electroweak sector and
for probing physics beyond the Standard Model through vector boson
scattering processes.

\begin{acknowledgments}
 The numerical calculations reported in this paper were partially performed at TUBITAK ULAKBIM, High Performance and Grid Computing Center (TRUBA resources).
\end{acknowledgments}

\end{document}